\documentclass[twocolumn,prl,preprintnumbers,superscriptaddress,amsmath,amssymb,english]{revtex4}
\usepackage{amsmath,amssymb,amsfonts,mathrsfs}
\usepackage{amsthm}
\usepackage{CJK}
\usepackage{bm}
\usepackage{lipsum}
\usepackage{babel}
\usepackage{graphicx}
\allowdisplaybreaks
\usepackage{makecell}
\usepackage{booktabs}
\usepackage{multirow}
\usepackage{gensymb}
\usepackage{rotating}
\usepackage{float}
\usepackage{braket}
\usepackage[breaklinks]{hyperref}
\usepackage{color}
\hypersetup{colorlinks=true, linkcolor=blue, citecolor=blue, filecolor=blue, urlcolor=blue}

\begin{document}

\title{Floquet Valley-Polarized Quantum Anomalous Hall State in Nonmagnetic Heterobilayers}

\author{Fangyang Zhan}
\affiliation{Institute for Structure and Function $\&$ Department of Physics, Chongqing University, Chongqing 400044, P. R. China}
\affiliation{Chongqing Key Laboratory for Strongly Coupled Physics, Chongqing 400044, P. R. China}

\author{Zhen Ning}
\affiliation{Institute for Structure and Function $\&$ Department of Physics, Chongqing University, Chongqing 400044, P. R. China}
\affiliation{Chongqing Key Laboratory for Strongly Coupled Physics, Chongqing 400044, P. R. China}

\author{Li-Yong Gan}

\affiliation{Institute for Structure and Function $\&$ Department of Physics, Chongqing University, Chongqing 400044, P. R. China}
\affiliation{Chongqing Key Laboratory for Strongly Coupled Physics, Chongqing 400044, P. R. China}
\affiliation{Center for Computational Science and Engineering, Southern University of Science and Technology, Shenzhen 518055, P. R. China}
\affiliation{Center for Quantum materials and devices, Chongqing University, Chongqing 400044, P. R. China}

\author{Baobing Zheng}
\affiliation{College of Physics and Optoelectronic Technology, Baoji University of Arts and Sciences, Baoji 721016, P. R. China}
\affiliation{Institute for Structure and Function $\&$ Department of Physics, Chongqing University, Chongqing 400044, P. R. China}
\affiliation{Chongqing Key Laboratory for Strongly Coupled Physics, Chongqing 400044, P. R. China}

\author{Jing Fan}
\affiliation{Center for Computational Science and Engineering, Southern University of Science and Technology, Shenzhen 518055, P. R. China}

\author{Rui Wang}
\email[]{rcwang@cqu.edu.cn}
\affiliation{Institute for Structure and Function $\&$ Department of Physics, Chongqing University, Chongqing 400044, P. R. China}
\affiliation{Chongqing Key Laboratory for Strongly Coupled Physics, Chongqing 400044, P. R. China}
\affiliation{Center for Computational Science and Engineering, Southern University of Science and Technology, Shenzhen 518055, P. R. China}
\affiliation{Center for Quantum materials and devices, Chongqing University, Chongqing 400044, P. R. China}

\begin{abstract}
The valley-polarized quantum anomalous Hall (VQAH) state, which forwards a strategy for combining valleytronics and spintronics with nontrivial topology, attracts intensive interest in condensed-matter physics. So far, the explored VQAH states have still been limited to magnetic systems. Here, using the low-energy effective model and Floquet theorem, we propose a different mechanism to realize the Floquet VQAH state in nonmagnetic heterobilayers under light irradiation. We then realize this proposal via first-principles calculations in transition metal dichalcogenide heterobilayers, which initially possess the time-reversal invariant valley quantum spin Hall (VQSH) state. By irradiating circularly polarized light, the time-reversal invariant VQSH state can evolve into the VQAH state, behaving as an optically switchable topological spin-valley filter. These findings not only offer a rational scheme to realize the VQAH state without magnetic orders, but also pave a fascinating path for designing topological spintronic and valleytronic devices.
\end{abstract}



\maketitle
The electronic band structures of a solid usually have local extrema of valence and/or conduction band. These local extrema are termed as valleys, which provide an additional degree of freedom to characterize low-energy excitations beyond charge and spin \cite{2007Valley,PhysRevLett.99.236809,PhysRevLett.108.196802,Xu2014}. Similar to spin polarization in spintronics, the valley freedom can also encode information via valley polarization, thus giving rise to emergence of valleytronics. Over the past decade, valleytronics has attracted intense interest of studying the two-dimensional (2D) graphene and transitional metal dichalcogenides (TMD) \cite{PhysRevB.77.235406,RevModPhys.82.1959,Cao2012,PhysRevX.5.011040,PhysRevLett.118.096602,PhysRevApplied.11.044033}, in which intervalley interactions between two inequivalent valleys $K$ and $K'$  of hexagonal Brillouin Zone (BZ) are nearly absent. The inequivalent valleys can generate fascinating valley-dependent phenomena, such as valley Hall effects \cite{Mak1489,Gorbachev448}, valley-dependent optical selection rules \cite{Zeng2012,Mak2012,Optical2013}, valley magnetic effects \cite{PhysRevB.92.121403,2014Magnetic,PhysRevLett.114.037401}, and beyond \cite{2013Magnetoelectric,PhysRevLett.110.197402,Schaibley2016,ISLAM2016304,PhysRevLett.125.157402,PhysRevLett.127.116402}.

Compared with graphene, TMD have competitive advantages for fundamental and applied levels in valleytronics due to their spontaneous inversion ($\mathcal{I}$) symmetry breaking and strong spin-orbital coupling (SOC) effects. The $\mathcal{I}$-symmetry breaking in TMD leads to local nonzero Berry curvature near $K$ and $K'$ points, allowing us to define a valley-resolved Chern number $\mathcal{C}_{\tau}$, where $\tau$ denotes the valley $K$ or $K'$ \cite{PhysRevLett.106.156801,Zhang10546,PhysRevB.88.161406}. Meanwhile, strong SOC effects combining with $\mathcal{I}$-symmetry breaking generate the unique spin-valley coupling effect \cite{PhysRevLett.108.196802}, resulting in $\mathcal{C}_{\tau}=\mathcal{C}^{s}$, where $\mathcal{C}^{s}$ ($s=\uparrow,\downarrow$) is the spin-resolved Chern number. Such locking of spin to valley is more intriguing when the time-reversal ($\mathcal{T}$) symmetry is further broken. The $\mathcal{T}$-symmetry breaking could lead to the attractive valley-polarized quantum anomalous Hall (VQAH) state, which carries dissipationless chiral edge channels characterized by nonzero integer Chern numbers $\mathcal{C}$ \cite{PhysRevB.88.161406,PhysRevLett.112.106802,PhysRevLett.119.046403,PhysRevB.98.201407,PhysRevB.91.045404}. The VQAH phase was first proposed by Pan \textit{et al.} through artificially tuning the Rashba SOC in silicene \cite{PhysRevLett.112.106802}, and afterwards was demonstrated in bilayer graphene systems \cite{PhysRevB.101.155425,PhysRevB.104.L161113} and magnetic mateirals crystallizing in hexagonal lattices \cite{PhysRevLett.119.046403,nanolett.5b01373,PhysRevB.91.165430,PhysRevB.91.041303,PhysRevB.94.235449,QianSui97301,WOSConcepts,PhysRevB.104.L121403}.

These advances mentioned above forward a strategy for designing valleytronic devices combining with nontrivial topology. While recent advancements have been very encouraging, realistic VQAH candidates predicted so far have still been limited to systems with magnetic orders. As is well known, the intrinsic magnetic order in a 2D system is often strongly suppressed by thermal fluctuations \cite{PhysRevLett.17.1133}, thus making their experimental studies difficult. 
As the VQAH state is one of the most interestingly topological phenomena, it is highly desirable to explore different mechanism for finding the feasible VQAH state in realistic materials.

\begin{figure}
    \centering
    \includegraphics[scale=1]{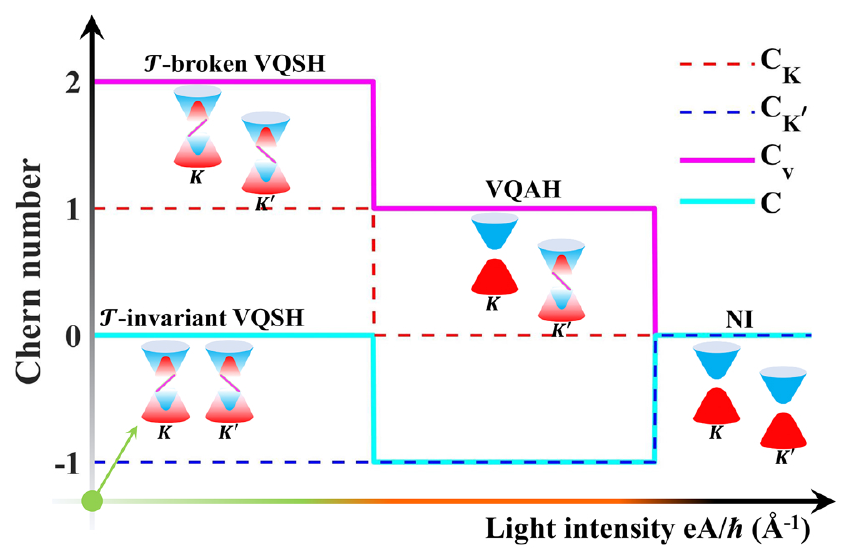}
    \caption{Schematic illustration of the evolution of topological phases in TMD heterobilayers under irradiation of CPL. The different topological phases are distinguished by  $\mathcal{C}=\mathcal{C}_K+\mathcal{C}_{K'}$ and valley Chern number $\mathcal{C}_v=\mathcal{C}_K-\mathcal{C}_{K'}$ with valley-resolved Chern number $\mathcal{C}_K$ and $\mathcal{C}_K'$.
    \label{FIG1}}
\end{figure}
Recently, Floquet-Bloch states driven by periodic light fields in condensed matter systems \cite{science.1239834,advs.202101508, PhysRevLett.116.016802,lindner2011floquet,PhysRevB.90.115423,PhysRevLett.110.026603,PhysRevLett.110.200403,PhysRevLett.117.087402,PhysRevB.97.155152,PhysRevLett.120.237403,hubener2017creating,mciver2020light} offer a different route to dynamically break the $\mathcal{T}$-symmetry, and provides a powerful path towards Floquet engineering topological states with high tunability \cite{pssr.201206451,annurev-conmatphys-031218-013423}.
In this work, we propose a new design scheme for realizing the VQAH state in realistic materials through introducing light fields. 
Based on the low-energy effective model and first-principles calculations, we show that TMD heterobilayers subject to circularly polarized light (CPL) can achieve this proposal.
Under irradiation of CPL, the initial $\mathcal{T}$-invariant VQSH state in TMD heterobilayers can evolve into the $\mathcal{T}$-broken VQSH state and then into the VQAH state, resulting in an optically switchable topological spin-valley filter.

To elucidate the light-modulated topological phases, we start from a $4\times4$ $\mathbf{k}\cdot \mathbf{p}$ model of heterobilayer formed by two massive Dirac materials with a specific stacking order, such as TMD \cite{2017Topological}. The Hamiltonian around the $K$ or $K'$ valley at the first order of $k$ can be written as (see details in Supplemental Material (SM) \cite{SM})
\begin{equation}\label{Eq.1}
\begin{split}
H_{\tau}(\mathbf{k})&= \boldsymbol s_x
(\tau v_1k_{x}\boldsymbol\sigma_{x}+ v_2 k_{y}\boldsymbol\sigma_{y}) +
\boldsymbol s_y
( v_1k_{y}\boldsymbol\sigma_{x}- \tau v_2 k_{x}\boldsymbol\sigma_{y}) \\
& +\frac{\Delta}{2}\boldsymbol s_z\boldsymbol\sigma_{z} +M_2(\boldsymbol s_z\boldsymbol\sigma_{0}-\boldsymbol s_0\boldsymbol\sigma_{0}) +
  M_1(\boldsymbol s_0\boldsymbol\sigma_{z}-\boldsymbol s_z\boldsymbol\sigma_{z}) \\
&  + \lambda(\boldsymbol s_x\boldsymbol\sigma_{0} - \boldsymbol s_x\boldsymbol\sigma_{z}),
\end{split}
\end{equation}
in which $v_{1,2} = {(v_l \pm v_u)}/{2}$, $M_{1,2} = {(M_l \pm M_u)}/{4}$ with Fermi velocity $\nu_{u}(\nu_{l})$ and Dirac mass $M_{u}(M_{l})$ in the upper (lower) layer, $\lambda$ denotes interlayer hopping, $\tau=+ 1 (-1)$ indicates the $K$ ($K'$) valley. $\mathit{\Delta}=\mathit{\Delta}_{g}-U$ is the band gap of heterobilayer, where $U$ is the interlayer bias proportional to the perpendicular electric field, and $\mathit{\Delta}_{g}$ is the intrinsic band gap of heterobilayer. The gap-closing topological phase transition occurs when $\mathit{\Delta}$ varies from $\mathit{\Delta}>0$ to $\mathit{\Delta}<0$. Here we only focus on neighborhood regime around the critical point $\mathit{\Delta}\thicksim0$, i.e., $M_{u}(M_{l}) \gg \mathit{\Delta}$. In this case, the $4\times4$ Hamiltonian Eq.(\ref{Eq.1}) can be projected to a $2\times2$ Hamiltonian \cite{2017Topological,SM}.

Under irradiation of time-periodic and spatially homogeneous CPL with vector potential of $\mathbf{A}(t)=A[\cos(\omega t), \pm \eta \sin(\omega t)]$, where $\eta=+1 (-1)$ denotes left- (right-) handed CPL, $\omega$ is frequency, and $A$ is the amplitude, the periodic driving field is
introduced in the Hamiltonian by using the minimal coupling substitution $H_{\tau}(\mathbf{k},t)=H_{\tau}(\mathbf{k}+e\mathbf{A(t)})$~\cite{PhysRevA.27.72,PhysRevLett.110.200403}. The time-periodic Hamiltonian can be expanded as $H_{\tau}(\mathbf{k},t)=\sum_{m}H_{\tau}^{m}(\mathbf{k})e^{im\omega t}$. In the high frequency regime, we can obtain the effective time-independent Floquet Hamiltonian as \cite{SM,PhysRevX.4.031027}
\begin{equation}\label{Eq.2}
\begin{split}
H_{\tau}^{F}(\mathbf{k}) &= H^{0}_{\tau}(\mathbf{k})+\sum_{m\ge 1}\frac{\left[H_{\tau}^{-m},H_{\tau}^{m}\right]}{m\hbar \omega} \\
&= \tilde{d}_{0}(k)\sigma_{0}+\tilde{\mathbf{d}}_{\tau}(\mathbf{k}) \cdot \boldsymbol{\sigma}.
\end{split}
\end{equation}
Here, the light renormalized $\tilde{d}_{0}(k) = \tilde{D}-Dk^2$ and $\tilde{\mathbf{d}}_{\tau}(\mathbf{k})=(\tau \tilde{C}k_{x}, \tilde{C}k_{y}, \tilde{M}_{\tau}-Bk^{2})$, where  $\tilde{M}_{\tau}=\frac{\mathit{\Delta}}{2}+(\frac{eA}{\hbar})^{2}[\frac{P}{2}+\frac{8\eta\tau}{\hbar \omega}(\frac{\nu_{l}}{M_{l}})^{2}\lambda]$, $B=-\frac{P}{2}$, $\mathit{P}=\nu^{2}_{u}/M_{u}+\nu^{2}_{l}/M_{l}$, and other parameters are given in SM \cite{SM}. The valley-resolved Chern number can be analytically obtained as $C_{\tau}=-\frac{\tau}{2}[\mathrm{sgn}(\tilde{M_{\tau}}) + \mathrm{sgn}(B)]$.
Due to the locking of spin to valley, two inequivalent valleys must belong to opposite spin. Without loss of generality, we can define the $K$ valley with spin-up states and the $K'$ valley with spin-down states. The valley-related topological phases can be characterized by two nonzero invariants, such as the Chern number $\mathcal{C}=\mathcal{C}_K+\mathcal{C}_{K'}$ and valley (or spin) Chern number $\mathcal{C}_v=\mathcal{C}_K-\mathcal{C}_{K'}$. The evolution of topological phases under irradiation of CPL is then established according to topological invariants. In the main text, we only present results subject to left-handed CPL, which polarizes in the $x-y$ plane and propagates along the stacking direction ($-z$) of the heterobilayer. The results under right-handed CPL only switch the states between two valleys, included in the SM \cite{SM}.

The renormalized Dirac mass $\tilde{M}_{\tau}$ always monotonously increases with the increasing light intensity of CPL [see Fig. S1 in the SM \cite{SM}]. Thereby, the topological phase transition requires that the proposed heterobilayer must initially be in the band inverted regime (i.e., $\mathit{\Delta}<0$). In such case, band topology for each valley can be considered as a spin-Chern insulator. Without irradiation of CPL, the $\mathcal{T}$-symmetry guarantees two spin-Chern insulators with opposite chirality at the $K$ and $K'$ valleys, and thus a $\mathcal{T}$-invariant valley quantum spin Hall (VQSH) phase with two Kramers-degenerate nontrivial edge channels is formed. The evolution of topological phases with the increase of light intensity is illustrated in Fig. \ref{FIG1}, together with the calculated valley-resolved Chern number, Chern number, and valley Chern number.
Clearly, the light-induced $\mathcal{T}$-symmetry breaking makes that different valley regimes exhibit different responses to CPL. With a finite light intensity, there is no band gap closing at two valleys and thus two spin states have different inverted band gaps, leading to a $\mathcal{T}$-broken VQSH state \cite{PhysRevLett.107.066602}. The $\mathcal{T}$-invariant and $\mathcal{T}$-broken VQSH states are equivalent topological phases characterized by the same valley (or spin) Chern number $\mathcal{C}_v$. With increasing the light intensity, the spin-up states at the $K$ valley can undergo a topological phase transition with band gap closing and reopening, while the spin-down states at the $K'$ valley always preserve its nontrivial topology. As a result, the VQAH phase characterized by nonzero Chern number $\mathcal{C}$ is present.

\begin{figure}
    \centering
    \includegraphics[scale=1]{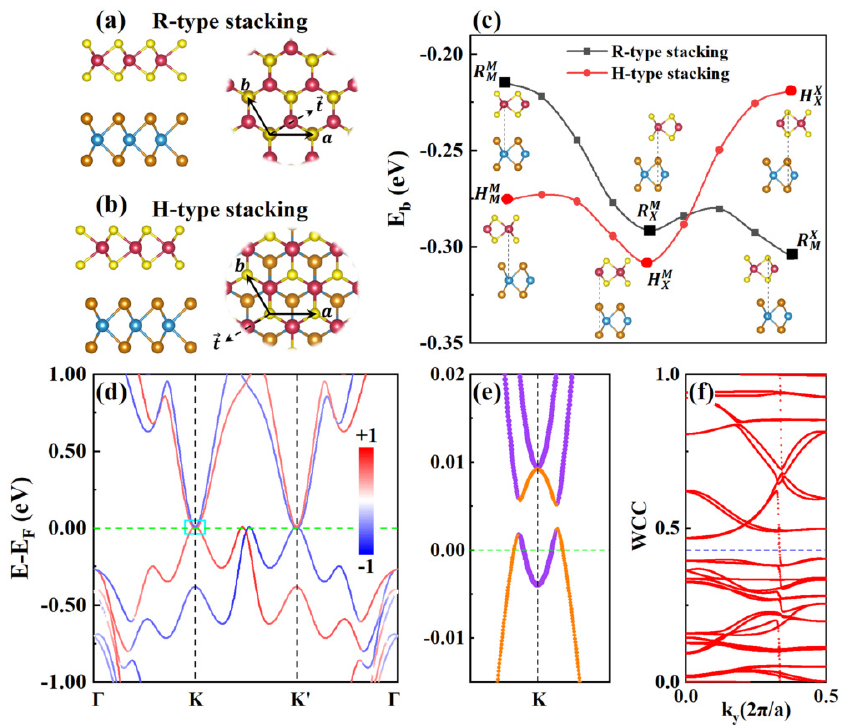}
    \caption{Two types of stacking orientations of TMD heterobilayers: (a) The R-type stacking (two layers have same orientations) and (b) The H-type stacking (two layers have opposite orientations). (c) The calculated binding energy $E_b$ of WS$_2$/WTe$_2$ heterobilayer as a function of stacking fault vector $\mathbf{t}$, which parallels the dashed-line arrow in the panel (a) or panel (b). The high-symmetry stacking configurations labeled as $H(R)_l^u$, which denotes a H(R)-type stacking with either the metal ($M$) or chalcogen ($X$) sites of upper layer are vertically aligned with the $M$ or $X$ sites of lower layer. (d) The spin-resolved band structure of $H_X^M$ stacking WS$_2$/WTe$_2$ heterobilayer in the presence of SOC. The red and blue colors indicate the $z$-component of spin-up and spin-down states, respectively. (e) The orbital-resolved bands around the $K$ valley. The component of W-$d_{z^{2}}$ orbital of upper WS$_2$ layer (W-$d_{xy}$\&$d_{x^{2}-{y^{2}}}$ orbital of lower WTe$_2$ layer) is proportional to the width of the purple (orange) curve. A perpendicular electric field is fixed at -0.02 V/\AA. (f) The Wannier charge centers (WCCs) confirms the VQSH phase.
     \label{FIG2}}
\end{figure}

To further confirm topological phenomena realized in TMD heterobilayers under the irradiation of CPL, we carried out first-principles calculations based on the density-functional theory \cite{PhysRev.140.A1133} as implemented in the Vienna ab initio simulation package \cite{PhysRevB.54.11169}. By projecting plane waves onto localized Wannier basis \cite{Mostofi2014}, we studied the photon-dressed band structures employing the Wannier tight-binding (TB) Hamiltonian combining with the Floquet theorem \cite{PhysRevB.102.201105}. The calculated details are included in the SM \cite{SM}. Here, without loss of generality, we start from a series of potential TMD heterobilayers of MX$_2$/WTe$_2$ (M=Mo, W; X= S, Se). The choice of these candidates is based on their work function differences between the upper and lower layers. The calculated results indicate that band gaps of heterobilayers decrease with the increase of work function differences (see Table SI and Fig. S2 in the SM \cite{SM}). To obtain band inversion in the presence of SOC, the large work function difference of two layers is preferred. As expected, it is found that initial band inversion can occur in the WS$_2$/WTe$_2$ and MoS$_2$/WTe$_2$ heterobilayers (see Fig. S2 in the SM \cite{SM}).
Here, we use the  WS$_2$/WTe$_2$ heterobilayer as an example to illustrate the light-driven topological phase transition. The results of MoS$_2$/WTe$_2$ heterobilayer are given in the SM \cite{SM}.

As shown in Figs. \ref{FIG2}(a)-\ref{FIG2}(c), there are three high-symmetry configurations of H-type (R-type) stacking associated with the $C_3$ rotational symmetry \cite{2019Evidence,2019Signatures}, labeled as $H(R)_l^u$ that is the H(R)-type stacking with either the metal ($M$) or chalcogen ($X$) sites of upper layer are vertically aligned with the $M$ or $X$ sites of lower layer.
Starting from the $H_M^M$ (or $R_M^M$) configuration, the calculated binding energy $E_b$ of WS$_2$/WTe$_2$ heterobilayer as a function of stacking fault vector $\mathbf{t}$ is plotted in Fig. \ref{FIG2}(c).
The results show that the H-type stacking $H_X^M$ corresponds to the global minima,
and thus this configuration of WS$_2$/WTe$_2$ heterobilayer has the highest thermodynamic stability in the WS$_2$/WTe$_2$ heterobilayer.


As illustrated in Fig. \ref{FIG2}(d), the spin-resolved band structure within SOC exhibits large spin-orbital splitting. We can clearly see that valence band regimes around the $K$ and $K'$ valleys are respectively contributed by spin-up and spin-down states, revealing the locking effect of spin to valley in the WS$_2$/WTe$_2$ heterobilayer.
We depict orbital-resolved band structures near the $K$ valley in Fig. \ref{FIG2}(e). The bands show that W-$d_{z^{2}}$ orbital of upper WS$_2$ layer and W-$d_{xy}$\&$d_{x^{2}-{y^{2}}}$ orbitals of lower WTe$_2$ layer are inverted at the $K$ valley, indicating the $H_X^M$ stacking WS$_2$/WTe$_2$ heterobilayer hosts the inverted type-II band alignment. Because two valleys are correlated by the $\mathcal{T}$-symmetry, the spin-down states at the $K'$ valley also possesses the same band inversion. That is to say, a $\mathcal{T}$-invariant VQSH state is present in the $H_X^M$ stacking WS$_2$/WTe$_2$ heterobilayer. The band topology is further verified by the evolution of Wannier charge centers (WCCs)  based on the Wilson loop method \cite{PhysRevB.84.075119}, as shown in Fig. \ref{FIG2}(f). Besides, it is worth noting that the presence of topological phase is strongly dependent on the stacking order in TMD heterobilayers \cite{2017Topological}. For the H-type WS$_2$/WTe$_2$ heterobilayer, only the $H_X^M$ stacking is topologically nontrivial and the $H_M^M$ and $H_X^X$ stacking are topologically trivial (see Fig. S4 in the SM \cite{SM}).

\begin{figure}
    \centering
    \includegraphics{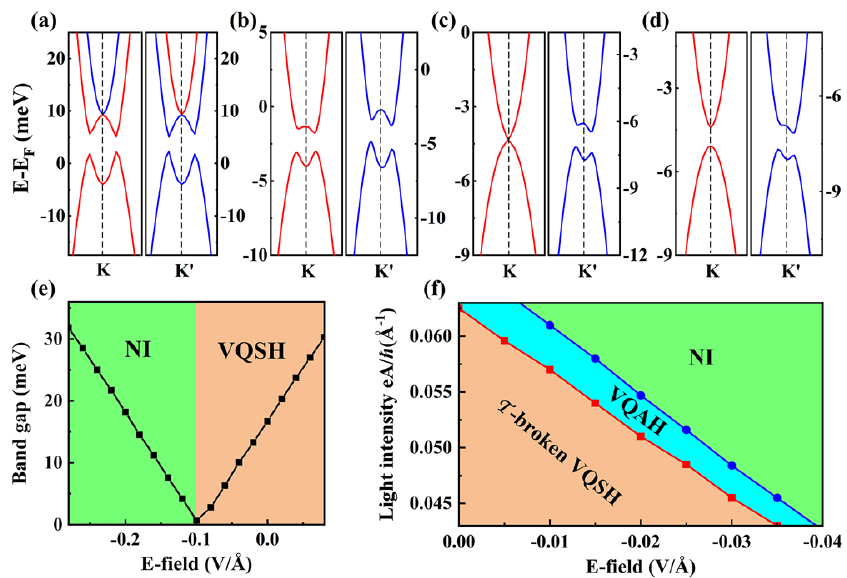}
    \caption{(a)-(d) The evolution of spin-resolved band structures of $H_X^M$ stacking WS$_2$/WTe$_2$ heterobilayer around the $K$ and $K'$ valleys with a light intensity $eA/\hbar$ of 0, 0.045, 0.051, and 0.053 {\AA}$^{-1}$.  The red and blue colors indicate the $z$-component of spin-up and spin-down states, respectively. A perpendicular electric field is fixed at -0.02 V/\AA. (e) Band gap at $K$ point as a function of the electric field strength without light irradiation. (f) The phase diagram dependent on electric field strength and light intensity.
    \label{FIG3}}
\end{figure}

After depicting the band topology of $H_X^M$ stacking WS$_2$/WTe$_2$ heterobilayer, we study the evolution of topological phases subject to light irradiation of left-handed CPL.
Once light irradiation of CPL is applied, the broken $\mathcal{T}$-symmetry will lift the Kramers degeneracy of two inequivalent valleys, and bands at the $K$ and $K'$ valleys will exhibit different behaviors. In Figs. \ref{FIG3}(a)-\ref{FIG3}(d), we illustrate the evolution of spin-resolved band structures around the $K$ and $K'$ valleys, respectively. As expected, one can find that the bands of spin-up states around the $K$ valley has been more drastically modified than those of spin-down states around the $K'$ valley. More importantly, with increasing light intensity, the band gap of spin-up states first closes and then reopens; that is, only spin-down states preserve the inverted band topology, resulting in a VQSH-to-VQAH topological phase transition.
As shown in Fig. \ref{FIG3}(c), the critical gapless point of spin-up states occurs at $eA_c/\hbar = $0.051 {\AA}$^{-1}$ (corresponding to the electric field strength of $2.553\times10^9$ V/m or peak intensity of $8.576\times10^{11}$ $\mathrm{W/cm}^2$; this light intensity can be easily realized in experiments).
When the light intensity is $0<eA/\hbar<eA_c/\hbar$, the bands around two valleys have different inverted band gaps as shown in Fig. \ref{FIG3}(b), indicating the presence of the $\mathcal{T}$-broken VQSH phase.

The $H_X^M$ stacking WS$_2$/WTe$_2$ heterobilayer belongs to type-II band alignment, so the interlayer bias can change its band gap and result in a band gap-closing topological phase transition [Eq. (\ref{Eq.1})] \cite{2017Topological}. As shown in Fig. \ref{FIG3}(e), we can see that an external perpendicular electric field can force the VQSH state to a normal insulating (NI) state in the absence of light irradiation. To fully understand cooperative effects of static electric fields and dynamic light fields, we carry out a comprehensive study on topological phases as functions of electric field strength and light intensity of CPL as illustrated in Fig. \ref{FIG3}(f). It is found that the perpendicular downward electric field can effectively reduce the threshold of light intensity of the VQSH-to-VQAH topological phase transition, thus facilitating to achieve the Floquet VQAH state in experiments. 

\begin{figure}
    \centering
    \includegraphics[scale=1.0]{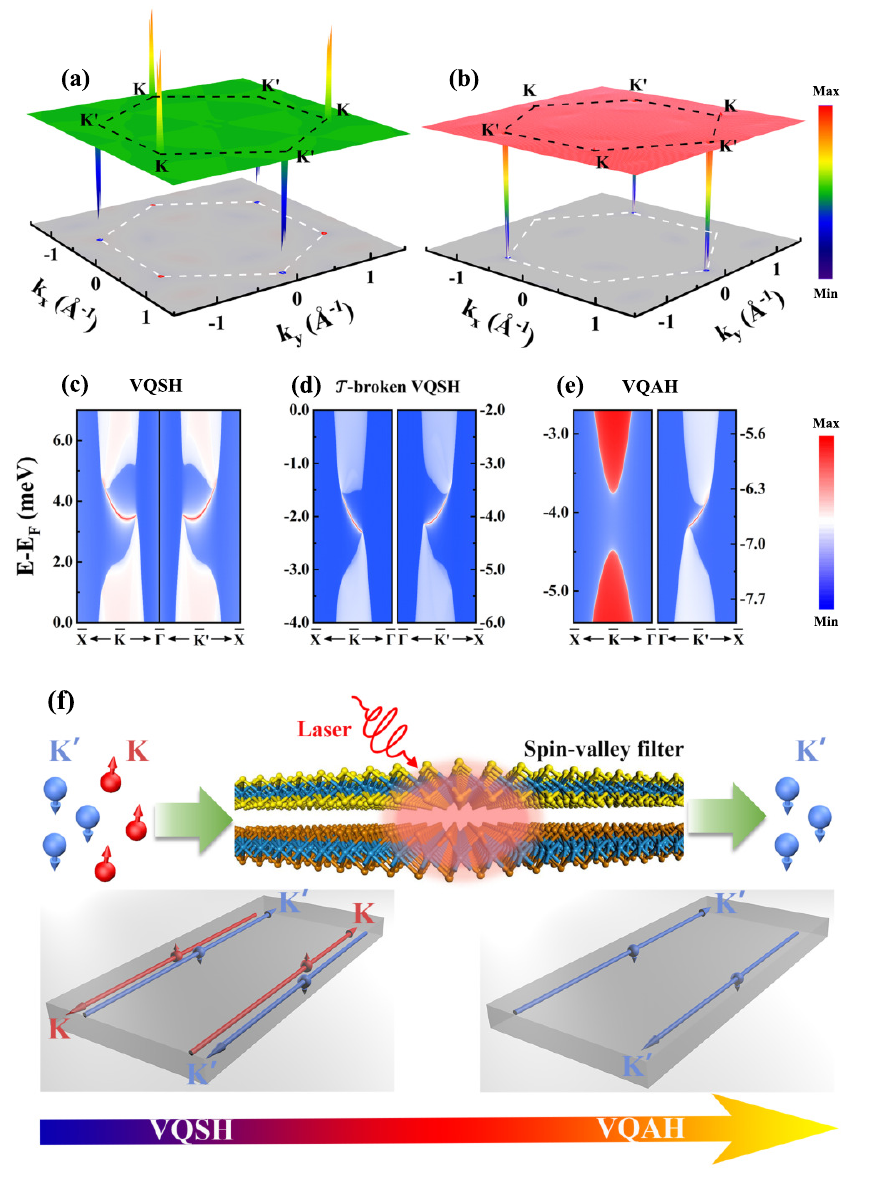}
    \caption{ The distribution of the Berry curvature $\Omega_z(\mathbf{k})$ for (a) $\mathcal{T}$-broken (or invariant) VQSH state and (b) VQAH state in the $k_x-k_y$ plane, respectively. The hexagonal Brilllouin zone is marked by dashed lines. The calculated LDOS projected on a semi-infinite ribbon of zigzag edge with a light intensity $eA/\hbar$ of (c) 0.0, (d) 0.045, and (e) 0.053 {\AA}$^{-1}$, which respectively corresponds to $\mathcal{T}$-invariant VQSH state,  $\mathcal{T}$-broken VQSH state, and VQAH state. A perpendicular electric field is fixed at -0.02 V/\AA. (f) Schematic illustration of an optically switchable topological spin-valley filter of TMD  heterobilayers.
    \label{FIG4}}
\end{figure}

The Berry curvature is an important quantity to characterize specific topological phases. As shown in Figs. \ref{FIG4}(a) and \ref{FIG4}(b), we show the distribution of Berry curvature $\Omega_z(\mathbf{k})$ for the $\mathcal{T}$-broken (or invariant) VQSH and VQAH phases in the $k_x-k_y$ plane, respectively. For the $\mathcal{T}$-broken (or invariant) VQSH phase, the Berry curvature $\Omega_z$ with opposite values diverges at the $K$ and $K'$ valleys and rapidly vanishes away from the $K$ and $K'$ valleys, thus allowing to well define the valley-resolved Chern number $\mathcal{C}_{\tau}$. By integrating the Berry curvature, we can obtain $\mathcal{C}_K=1$ and $\mathcal{C}_{K'}=-1$ for the $\mathcal{T}$-broken (or invariant) VQSH state, and the valley Chern number is $\mathcal{C}_v= 2$. For the VQAH state, the Berry curvature $\Omega_z$ only distributes and diverges near $K'$ valleys, giving $\mathcal{C}_K=0$ and $\mathcal{C}_{K'}=-1$, and then the Chern number is $\mathcal{C}=-1$. The numerical results from first-principles calculations match well with those from low-energy effective model as shown in Fig. \ref{FIG1}.

The topological phase with a specific topological invariant gives rise to uniquely nontrivial edge states. To directly illustrate this, we calculate the local density of states (LDOS) based on the iterative Green's method \cite{Sancho_1985,WU2017} using the Floquet Wannier TB Hamiltonian \cite{SM}. The calculated LDOS projected on a semi-infinite ribbon of $H_X^M$ stacking WS$_2$/WTe$_2$ heterobilayer with a light intensity $eA/\hbar$ of 0, 0.045, and 0.053 {\AA}$^{-1}$  are plotted in Figs. \ref{FIG4}(c)-\ref{FIG4}(e), respectively. Without light irradiation, the $\mathcal{T}$-invariant VQSH state shows that two opposite chiral edge states with Kramers degeneracy are visible at the $K$ and $K'$ valleys [see Fig. \ref{FIG4}(c)]. With increasing light intensity to 0.045 {\AA}$^{-1}$, the $\mathcal{T}$-broken VQSH state removes the Kramers degeneracy and exhibits different inverted band gaps at the $K$ and $K'$ valleys, but the chiral edge states are also visibly present [see Fig. \ref{FIG4}(d)], confirming its nontrivial feature. As shown in Fig. \ref{FIG4}(e), the VQAH state possesses one chiral edge state connecting the valence and conduction bands around the $K'$ valley, while bands around the $K$ valley exhibit the topological trivial NI state. Considering the locking of spin to valley in TMD, the VQAH state with one spin- and valley-resolved chiral edge channel in our proposed TMD heterobilayers under light irradiation behaves as a perfect topological spin-valley filter [see Fig. \ref{FIG4}(f)].

In summary, we proposed a rational principle to achieve the VQAH state in nonmagnetic heterobilayers of TMD by introducing light irradiation of CPL. From a $\mathcal{T}$-invariant VQSH state, the $\mathcal{T}$-symmetry breaking induced by CPL leads to different responses around two valley regimes, which guarantees that the topological phase transition associated with band gap closing and reopening only occurs at one valley. As a result, one valley remains the spin-Chern insulating state, while the other valley is the NI state. Furthermore, the VQAH and NI states around two valleys can be flexibly switched by left- or right-handed CPL \cite{SM}. Hence, our proposal not only offers a reliable scheme to realize the optically switchable topological spin-valley filter, but also opens a route towards designing topological VQAH-based devices controlled by light fields. In addition, our work purports to forward a strategy to realize VQAH states in nonmagnetic systems subjected to dynamical light fields, and is expected to obtain larger band gap VQAH effects in other valley-based materials in futures.

This work was supported by the National Natural Science Foundation of China (NSFC, Grants No. 11974062, No. 12047564, and No. 11704177), the Chongqing Natural Science Foundation (Grants No. cstc2019jcyj-msxmX0563), the Fundamental Research Funds for the Central Universities of China (Grant No. 2020CDJQY-A057), and the Beijing National Laboratory for Condensed Matter Physics.


\begin{thebibliography}{67}%
\makeatletter
\providecommand \@ifxundefined [1]{%
 \@ifx{#1\undefined}
}%
\providecommand \@ifnum [1]{%
 \ifnum #1\expandafter \@firstoftwo
 \else \expandafter \@secondoftwo
 \fi
}%
\providecommand \@ifx [1]{%
 \ifx #1\expandafter \@firstoftwo
 \else \expandafter \@secondoftwo
 \fi
}%
\providecommand \natexlab [1]{#1}%
\providecommand \enquote  [1]{``#1''}%
\providecommand \bibnamefont  [1]{#1}%
\providecommand \bibfnamefont [1]{#1}%
\providecommand \citenamefont [1]{#1}%
\providecommand \href@noop [0]{\@secondoftwo}%
\providecommand \href [0]{\begingroup \@sanitize@url \@href}%
\providecommand \@href[1]{\@@startlink{#1}\@@href}%
\providecommand \@@href[1]{\endgroup#1\@@endlink}%
\providecommand \@sanitize@url [0]{\catcode `\\12\catcode `\$12\catcode
  `\&12\catcode `\#12\catcode `\^12\catcode `\_12\catcode `\%12\relax}%
\providecommand \@@startlink[1]{}%
\providecommand \@@endlink[0]{}%
\providecommand \url  [0]{\begingroup\@sanitize@url \@url }%
\providecommand \@url [1]{\endgroup\@href {#1}{\urlprefix }}%
\providecommand \urlprefix  [0]{URL }%
\providecommand \Eprint [0]{\href }%
\providecommand \doibase [0]{https://doi.org/}%
\providecommand \selectlanguage [0]{\@gobble}%
\providecommand \bibinfo  [0]{\@secondoftwo}%
\providecommand \bibfield  [0]{\@secondoftwo}%
\providecommand \translation [1]{[#1]}%
\providecommand \BibitemOpen [0]{}%
\providecommand \bibitemStop [0]{}%
\providecommand \bibitemNoStop [0]{.\EOS\space}%
\providecommand \EOS [0]{\spacefactor3000\relax}%
\providecommand \BibitemShut  [1]{\csname bibitem#1\endcsname}%
\let\auto@bib@innerbib\@empty
\bibitem [{\citenamefont {Rycerz}\ \emph {et~al.}(2007)\citenamefont {Rycerz},
  \citenamefont {Tworzyd?o},\ and\ \citenamefont {Beenakker}}]{2007Valley}%
  \BibitemOpen
  \bibfield  {author} {\bibinfo {author} {\bibfnamefont {A.}~\bibnamefont
  {Rycerz}}, \bibinfo {author} {\bibfnamefont {J.}~\bibnamefont {Tworzyd?o}},\
  and\ \bibinfo {author} {\bibfnamefont {C.~W.~J.}\ \bibnamefont {Beenakker}},\
  }\href {https://doi.org/10.1038/nphys547} {\bibfield  {journal} {\bibinfo
  {journal} {Nat. Phys.}\ }\textbf {\bibinfo {volume} {3}},\ \bibinfo {pages}
  {172-175} (\bibinfo {year} {2007})}\BibitemShut {NoStop}%
\bibitem [{\citenamefont {Xiao}\ \emph {et~al.}(2007)\citenamefont {Xiao},
  \citenamefont {Yao},\ and\ \citenamefont {Niu}}]{PhysRevLett.99.236809}%
  \BibitemOpen
  \bibfield  {author} {\bibinfo {author} {\bibfnamefont {D.}~\bibnamefont
  {Xiao}}, \bibinfo {author} {\bibfnamefont {W.}~\bibnamefont {Yao}},\ and\
  \bibinfo {author} {\bibfnamefont {Q.}~\bibnamefont {Niu}},\ }\href
  {https://doi.org/10.1103/PhysRevLett.99.236809} {\bibfield  {journal}
  {\bibinfo  {journal} {Phys. Rev. Lett.}\ }\textbf {\bibinfo {volume} {99}},\
  \bibinfo {pages} {236809} (\bibinfo {year} {2007})}\BibitemShut {NoStop}%
\bibitem [{\citenamefont {Xiao}\ \emph {et~al.}(2012)\citenamefont {Xiao},
  \citenamefont {Liu}, \citenamefont {Feng}, \citenamefont {Xu},\ and\
  \citenamefont {Yao}}]{PhysRevLett.108.196802}%
  \BibitemOpen
  \bibfield  {author} {\bibinfo {author} {\bibfnamefont {D.}~\bibnamefont
  {Xiao}}, \bibinfo {author} {\bibfnamefont {G.-B.}\ \bibnamefont {Liu}},
  \bibinfo {author} {\bibfnamefont {W.}~\bibnamefont {Feng}}, \bibinfo {author}
  {\bibfnamefont {X.}~\bibnamefont {Xu}},\ and\ \bibinfo {author}
  {\bibfnamefont {W.}~\bibnamefont {Yao}},\ }\href
  {https://doi.org/10.1103/PhysRevLett.108.196802} {\bibfield  {journal}
  {\bibinfo  {journal} {Phys. Rev. Lett.}\ }\textbf {\bibinfo {volume} {108}},\
  \bibinfo {pages} {196802} (\bibinfo {year} {2012})}\BibitemShut {NoStop}%
\bibitem [{\citenamefont {Xu}\ \emph {et~al.}(2014)\citenamefont {Xu},
  \citenamefont {Yao}, \citenamefont {Xiao},\ and\ \citenamefont
  {Heinz}}]{Xu2014}%
  \BibitemOpen
  \bibfield  {author} {\bibinfo {author} {\bibfnamefont {X.}~\bibnamefont
  {Xu}}, \bibinfo {author} {\bibfnamefont {W.}~\bibnamefont {Yao}}, \bibinfo
  {author} {\bibfnamefont {D.}~\bibnamefont {Xiao}},\ and\ \bibinfo {author}
  {\bibfnamefont {T.~F.}\ \bibnamefont {Heinz}},\ }\href
  {https://doi.org/10.1038/nphys2942} {\bibfield  {journal} {\bibinfo
  {journal} {Nat. Phys.}\ }\textbf {\bibinfo {volume} {10}},\ \bibinfo {pages}
  {343} (\bibinfo {year} {2014})}\BibitemShut {NoStop}%
\bibitem [{\citenamefont {Yao}\ \emph {et~al.}(2008)\citenamefont {Yao},
  \citenamefont {Xiao},\ and\ \citenamefont {Niu}}]{PhysRevB.77.235406}%
  \BibitemOpen
  \bibfield  {author} {\bibinfo {author} {\bibfnamefont {W.}~\bibnamefont
  {Yao}}, \bibinfo {author} {\bibfnamefont {D.}~\bibnamefont {Xiao}},\ and\
  \bibinfo {author} {\bibfnamefont {Q.}~\bibnamefont {Niu}},\ }\href
  {https://doi.org/10.1103/PhysRevB.77.235406} {\bibfield  {journal} {\bibinfo
  {journal} {Phys. Rev. B}\ }\textbf {\bibinfo {volume} {77}},\ \bibinfo
  {pages} {235406} (\bibinfo {year} {2008})}\BibitemShut {NoStop}%
\bibitem [{\citenamefont {Xiao}\ \emph {et~al.}(2010)\citenamefont {Xiao},
  \citenamefont {Chang},\ and\ \citenamefont {Niu}}]{RevModPhys.82.1959}%
  \BibitemOpen
  \bibfield  {author} {\bibinfo {author} {\bibfnamefont {D.}~\bibnamefont
  {Xiao}}, \bibinfo {author} {\bibfnamefont {M.-C.}\ \bibnamefont {Chang}},\
  and\ \bibinfo {author} {\bibfnamefont {Q.}~\bibnamefont {Niu}},\ }\href
  {https://doi.org/10.1103/RevModPhys.82.1959} {\bibfield  {journal} {\bibinfo
  {journal} {Rev. Mod. Phys.}\ }\textbf {\bibinfo {volume} {82}},\ \bibinfo
  {pages} {1959} (\bibinfo {year} {2010})}\BibitemShut {NoStop}%
\bibitem [{\citenamefont {Cao}\ \emph {et~al.}(2012)\citenamefont {Cao},
  \citenamefont {Wang}, \citenamefont {Han}, \citenamefont {Ye}, \citenamefont
  {Zhu}, \citenamefont {Shi}, \citenamefont {Niu}, \citenamefont {Tan},
  \citenamefont {Wang}, \citenamefont {Liu},\ and\ \citenamefont
  {Feng}}]{Cao2012}%
  \BibitemOpen
  \bibfield  {author} {\bibinfo {author} {\bibfnamefont {T.}~\bibnamefont
  {Cao}}, \bibinfo {author} {\bibfnamefont {G.}~\bibnamefont {Wang}}, \bibinfo
  {author} {\bibfnamefont {W.}~\bibnamefont {Han}}, \bibinfo {author}
  {\bibfnamefont {H.}~\bibnamefont {Ye}}, \bibinfo {author} {\bibfnamefont
  {C.}~\bibnamefont {Zhu}}, \bibinfo {author} {\bibfnamefont {J.}~\bibnamefont
  {Shi}}, \bibinfo {author} {\bibfnamefont {Q.}~\bibnamefont {Niu}}, \bibinfo
  {author} {\bibfnamefont {P.}~\bibnamefont {Tan}}, \bibinfo {author}
  {\bibfnamefont {E.}~\bibnamefont {Wang}}, \bibinfo {author} {\bibfnamefont
  {B.}~\bibnamefont {Liu}},\ and\ \bibinfo {author} {\bibfnamefont
  {J.}~\bibnamefont {Feng}},\ }\href {https://doi.org/10.1038/ncomms1882}
  {\bibfield  {journal} {\bibinfo  {journal} {Nat. Commun.}\ }\textbf {\bibinfo
  {volume} {3}},\ \bibinfo {pages} {887} (\bibinfo {year} {2012})}\BibitemShut
  {NoStop}%
\bibitem [{\citenamefont {Marino}\ \emph {et~al.}(2015)\citenamefont {Marino},
  \citenamefont {Nascimento}, \citenamefont {Alves},\ and\ \citenamefont
  {Smith}}]{PhysRevX.5.011040}%
  \BibitemOpen
  \bibfield  {author} {\bibinfo {author} {\bibfnamefont {E.~C.}\ \bibnamefont
  {Marino}}, \bibinfo {author} {\bibfnamefont {L.~O.}\ \bibnamefont
  {Nascimento}}, \bibinfo {author} {\bibfnamefont {V.~S.}\ \bibnamefont
  {Alves}},\ and\ \bibinfo {author} {\bibfnamefont {C.~M.}\ \bibnamefont
  {Smith}},\ }\href {https://doi.org/10.1103/PhysRevX.5.011040} {\bibfield
  {journal} {\bibinfo  {journal} {Phys. Rev. X}\ }\textbf {\bibinfo {volume}
  {5}},\ \bibinfo {pages} {011040} (\bibinfo {year} {2015})}\BibitemShut
  {NoStop}%
\bibitem [{\citenamefont {An}\ \emph {et~al.}(2017)\citenamefont {An},
  \citenamefont {Xiao}, \citenamefont {Tu}, \citenamefont {Yu}, \citenamefont
  {Fal'ko},\ and\ \citenamefont {Yao}}]{PhysRevLett.118.096602}%
  \BibitemOpen
  \bibfield  {author} {\bibinfo {author} {\bibfnamefont {X.-T.}\ \bibnamefont
  {An}}, \bibinfo {author} {\bibfnamefont {J.}~\bibnamefont {Xiao}}, \bibinfo
  {author} {\bibfnamefont {M.~W.-Y.}\ \bibnamefont {Tu}}, \bibinfo {author}
  {\bibfnamefont {H.}~\bibnamefont {Yu}}, \bibinfo {author} {\bibfnamefont
  {V.~I.}\ \bibnamefont {Fal'ko}},\ and\ \bibinfo {author} {\bibfnamefont
  {W.}~\bibnamefont {Yao}},\ }\href
  {https://doi.org/10.1103/PhysRevLett.118.096602} {\bibfield  {journal}
  {\bibinfo  {journal} {Phys. Rev. Lett.}\ }\textbf {\bibinfo {volume} {118}},\
  \bibinfo {pages} {096602} (\bibinfo {year} {2017})}\BibitemShut {NoStop}%
\bibitem [{\citenamefont {Park}(2019)}]{PhysRevApplied.11.044033}%
  \BibitemOpen
  \bibfield  {author} {\bibinfo {author} {\bibfnamefont {C.}~\bibnamefont
  {Park}},\ }\href {https://doi.org/10.1103/PhysRevApplied.11.044033}
  {\bibfield  {journal} {\bibinfo  {journal} {Phys. Rev. Applied}\ }\textbf
  {\bibinfo {volume} {11}},\ \bibinfo {pages} {044033} (\bibinfo {year}
  {2019})}\BibitemShut {NoStop}%
\bibitem [{\citenamefont {Mak}\ \emph {et~al.}(2014)\citenamefont {Mak},
  \citenamefont {McGill}, \citenamefont {Park},\ and\ \citenamefont
  {McEuen}}]{Mak1489}%
  \BibitemOpen
  \bibfield  {author} {\bibinfo {author} {\bibfnamefont {K.~F.}\ \bibnamefont
  {Mak}}, \bibinfo {author} {\bibfnamefont {K.~L.}\ \bibnamefont {McGill}},
  \bibinfo {author} {\bibfnamefont {J.}~\bibnamefont {Park}},\ and\ \bibinfo
  {author} {\bibfnamefont {P.~L.}\ \bibnamefont {McEuen}},\ }\href
  {https://doi.org/10.1126/science.1250140} {\bibfield  {journal} {\bibinfo
  {journal} {Science}\ }\textbf {\bibinfo {volume} {344}},\ \bibinfo {pages}
  {1489} (\bibinfo {year} {2014})}\BibitemShut {NoStop}%
\bibitem [{\citenamefont {Gorbachev}\ \emph {et~al.}(2014)\citenamefont
  {Gorbachev}, \citenamefont {Song}, \citenamefont {Yu}, \citenamefont
  {Kretinin}, \citenamefont {Withers}, \citenamefont {Cao}, \citenamefont
  {Mishchenko}, \citenamefont {Grigorieva}, \citenamefont {Novoselov},
  \citenamefont {Levitov},\ and\ \citenamefont {Geim}}]{Gorbachev448}%
  \BibitemOpen
  \bibfield  {author} {\bibinfo {author} {\bibfnamefont {R.~V.}\ \bibnamefont
  {Gorbachev}}, \bibinfo {author} {\bibfnamefont {J.~C.~W.}\ \bibnamefont
  {Song}}, \bibinfo {author} {\bibfnamefont {G.~L.}\ \bibnamefont {Yu}},
  \bibinfo {author} {\bibfnamefont {A.~V.}\ \bibnamefont {Kretinin}}, \bibinfo
  {author} {\bibfnamefont {F.}~\bibnamefont {Withers}}, \bibinfo {author}
  {\bibfnamefont {Y.}~\bibnamefont {Cao}}, \bibinfo {author} {\bibfnamefont
  {A.}~\bibnamefont {Mishchenko}}, \bibinfo {author} {\bibfnamefont {I.~V.}\
  \bibnamefont {Grigorieva}}, \bibinfo {author} {\bibfnamefont {K.~S.}\
  \bibnamefont {Novoselov}}, \bibinfo {author} {\bibfnamefont {L.~S.}\
  \bibnamefont {Levitov}},\ and\ \bibinfo {author} {\bibfnamefont {A.~K.}\
  \bibnamefont {Geim}},\ }\href {https://doi.org/10.1126/science.1254966}
  {\bibfield  {journal} {\bibinfo  {journal} {Science}\ }\textbf {\bibinfo
  {volume} {346}},\ \bibinfo {pages} {448} (\bibinfo {year}
  {2014})}\BibitemShut {NoStop}%
\bibitem [{\citenamefont {Zeng}\ \emph {et~al.}(2012)\citenamefont {Zeng},
  \citenamefont {Dai}, \citenamefont {Yao}, \citenamefont {Xiao},\ and\
  \citenamefont {Cui}}]{Zeng2012}%
  \BibitemOpen
  \bibfield  {author} {\bibinfo {author} {\bibfnamefont {H.}~\bibnamefont
  {Zeng}}, \bibinfo {author} {\bibfnamefont {J.}~\bibnamefont {Dai}}, \bibinfo
  {author} {\bibfnamefont {W.}~\bibnamefont {Yao}}, \bibinfo {author}
  {\bibfnamefont {D.}~\bibnamefont {Xiao}},\ and\ \bibinfo {author}
  {\bibfnamefont {X.}~\bibnamefont {Cui}},\ }\href
  {https://doi.org/10.1038/nnano.2012.95} {\bibfield  {journal} {\bibinfo
  {journal} {Nat. Nanotech.}\ }\textbf {\bibinfo {volume} {7}},\ \bibinfo
  {pages} {490} (\bibinfo {year} {2012})}\BibitemShut {NoStop}%
\bibitem [{\citenamefont {Mak}\ \emph {et~al.}(2012)\citenamefont {Mak},
  \citenamefont {He}, \citenamefont {Shan},\ and\ \citenamefont
  {Heinz}}]{Mak2012}%
  \BibitemOpen
  \bibfield  {author} {\bibinfo {author} {\bibfnamefont {K.~F.}\ \bibnamefont
  {Mak}}, \bibinfo {author} {\bibfnamefont {K.}~\bibnamefont {He}}, \bibinfo
  {author} {\bibfnamefont {J.}~\bibnamefont {Shan}},\ and\ \bibinfo {author}
  {\bibfnamefont {T.~F.}\ \bibnamefont {Heinz}},\ }\href
  {https://doi.org/10.1038/nnano.2012.96} {\bibfield  {journal} {\bibinfo
  {journal} {Nat. Nanotech.}\ }\textbf {\bibinfo {volume} {7}},\ \bibinfo
  {pages} {494} (\bibinfo {year} {2012})}\BibitemShut {NoStop}%
\bibitem [{\citenamefont {Jones}\ \emph {et~al.}(2013)\citenamefont {Jones},
  \citenamefont {Yu}, \citenamefont {Ghimire}, \citenamefont {Wu},
  \citenamefont {Aivazian}, \citenamefont {Ross}, \citenamefont {Zhao},
  \citenamefont {Yan}, \citenamefont {Mandrus}, \citenamefont {Xiao},
  \citenamefont {Yao},\ and\ \citenamefont {Xu}}]{Optical2013}%
  \BibitemOpen
  \bibfield  {author} {\bibinfo {author} {\bibfnamefont {A.~M.}\ \bibnamefont
  {Jones}}, \bibinfo {author} {\bibfnamefont {H.~Y.}\ \bibnamefont {Yu}},
  \bibinfo {author} {\bibfnamefont {N.~J.}\ \bibnamefont {Ghimire}}, \bibinfo
  {author} {\bibfnamefont {S.~F.}\ \bibnamefont {Wu}}, \bibinfo {author}
  {\bibfnamefont {G.}~\bibnamefont {Aivazian}}, \bibinfo {author}
  {\bibfnamefont {J.~S.}\ \bibnamefont {Ross}}, \bibinfo {author}
  {\bibfnamefont {B.}~\bibnamefont {Zhao}}, \bibinfo {author} {\bibfnamefont
  {J.~Q.}\ \bibnamefont {Yan}}, \bibinfo {author} {\bibfnamefont {D.~G.}\
  \bibnamefont {Mandrus}}, \bibinfo {author} {\bibfnamefont {D.}~\bibnamefont
  {Xiao}}, \bibinfo {author} {\bibfnamefont {W.}~\bibnamefont {Yao}},\ and\
  \bibinfo {author} {\bibfnamefont {X.~D.}\ \bibnamefont {Xu}},\ }\href
  {https://doi.org/10.1038/nnano.2013.151} {\bibfield  {journal} {\bibinfo
  {journal} {Nat. Nanotech.}\ }\textbf {\bibinfo {volume} {8}},\ \bibinfo
  {pages} {634-638} (\bibinfo {year} {2013})}\BibitemShut {NoStop}%
\bibitem [{\citenamefont {Qi}\ \emph {et~al.}(2015)\citenamefont {Qi},
  \citenamefont {Li}, \citenamefont {Niu},\ and\ \citenamefont
  {Feng}}]{PhysRevB.92.121403}%
  \BibitemOpen
  \bibfield  {author} {\bibinfo {author} {\bibfnamefont {J.}~\bibnamefont
  {Qi}}, \bibinfo {author} {\bibfnamefont {X.}~\bibnamefont {Li}}, \bibinfo
  {author} {\bibfnamefont {Q.}~\bibnamefont {Niu}},\ and\ \bibinfo {author}
  {\bibfnamefont {J.}~\bibnamefont {Feng}},\ }\href
  {https://doi.org/10.1103/PhysRevB.92.121403} {\bibfield  {journal} {\bibinfo
  {journal} {Phys. Rev. B}\ }\textbf {\bibinfo {volume} {92}},\ \bibinfo
  {pages} {121403} (\bibinfo {year} {2015})}\BibitemShut {NoStop}%
\bibitem [{\citenamefont {Aivazian}\ \emph {et~al.}(2015)\citenamefont
  {Aivazian}, \citenamefont {Gong}, \citenamefont {Jones}, \citenamefont
  {Chu},\ and\ \citenamefont {Xu}}]{2014Magnetic}%
  \BibitemOpen
  \bibfield  {author} {\bibinfo {author} {\bibfnamefont {G.}~\bibnamefont
  {Aivazian}}, \bibinfo {author} {\bibfnamefont {Z.}~\bibnamefont {Gong}},
  \bibinfo {author} {\bibfnamefont {A.~M.}\ \bibnamefont {Jones}}, \bibinfo
  {author} {\bibfnamefont {R.~L.}\ \bibnamefont {Chu}},\ and\ \bibinfo {author}
  {\bibfnamefont {X.}~\bibnamefont {Xu}},\ }\href
  {https://doi.org/10.1038/nphys3201} {\bibfield  {journal} {\bibinfo
  {journal} {Nat. Phys.}\ }\textbf {\bibinfo {volume} {11}},\ \bibinfo {pages}
  {148} (\bibinfo {year} {2015})}\BibitemShut {NoStop}%
\bibitem [{\citenamefont {MacNeill}\ \emph {et~al.}(2015)\citenamefont
  {MacNeill}, \citenamefont {Heikes}, \citenamefont {Mak}, \citenamefont
  {Anderson}, \citenamefont {Korm\'anyos}, \citenamefont {Z\'olyomi},
  \citenamefont {Park},\ and\ \citenamefont {Ralph}}]{PhysRevLett.114.037401}%
  \BibitemOpen
  \bibfield  {author} {\bibinfo {author} {\bibfnamefont {D.}~\bibnamefont
  {MacNeill}}, \bibinfo {author} {\bibfnamefont {C.}~\bibnamefont {Heikes}},
  \bibinfo {author} {\bibfnamefont {K.~F.}\ \bibnamefont {Mak}}, \bibinfo
  {author} {\bibfnamefont {Z.}~\bibnamefont {Anderson}}, \bibinfo {author}
  {\bibfnamefont {A.}~\bibnamefont {Korm\'anyos}}, \bibinfo {author}
  {\bibfnamefont {V.}~\bibnamefont {Z\'olyomi}}, \bibinfo {author}
  {\bibfnamefont {J.}~\bibnamefont {Park}},\ and\ \bibinfo {author}
  {\bibfnamefont {D.~C.}\ \bibnamefont {Ralph}},\ }\href
  {https://doi.org/10.1103/PhysRevLett.114.037401} {\bibfield  {journal}
  {\bibinfo  {journal} {Phys. Rev. Lett.}\ }\textbf {\bibinfo {volume} {114}},\
  \bibinfo {pages} {037401} (\bibinfo {year} {2015})}\BibitemShut {NoStop}%
\bibitem [{\citenamefont {Gong}\ \emph {et~al.}(2013)\citenamefont {Gong},
  \citenamefont {Liu}, \citenamefont {Yu}, \citenamefont {Xiao}, \citenamefont
  {Cui}, \citenamefont {Xu},\ and\ \citenamefont {Yao}}]{2013Magnetoelectric}%
  \BibitemOpen
  \bibfield  {author} {\bibinfo {author} {\bibfnamefont {Z.}~\bibnamefont
  {Gong}}, \bibinfo {author} {\bibfnamefont {G.}~\bibnamefont {Liu}}, \bibinfo
  {author} {\bibfnamefont {H.}~\bibnamefont {Yu}}, \bibinfo {author}
  {\bibfnamefont {D.}~\bibnamefont {Xiao}}, \bibinfo {author} {\bibfnamefont
  {X.}~\bibnamefont {Cui}}, \bibinfo {author} {\bibfnamefont {X.}~\bibnamefont
  {Xu}},\ and\ \bibinfo {author} {\bibfnamefont {W.}~\bibnamefont {Yao}},\
  }\href {https://doi.org/10.1038/ncomms3053} {\bibfield  {journal} {\bibinfo
  {journal} {Nat. Commun.}\ }\textbf {\bibinfo {volume} {4}},\ \bibinfo {pages}
  {2053} (\bibinfo {year} {2013})}\BibitemShut {NoStop}%
\bibitem [{\citenamefont {Tabert}\ and\ \citenamefont
  {Nicol}(2013)}]{PhysRevLett.110.197402}%
  \BibitemOpen
  \bibfield  {author} {\bibinfo {author} {\bibfnamefont {C.~J.}\ \bibnamefont
  {Tabert}}\ and\ \bibinfo {author} {\bibfnamefont {E.~J.}\ \bibnamefont
  {Nicol}},\ }\href {https://doi.org/10.1103/PhysRevLett.110.197402} {\bibfield
   {journal} {\bibinfo  {journal} {Phys. Rev. Lett.}\ }\textbf {\bibinfo
  {volume} {110}},\ \bibinfo {pages} {197402} (\bibinfo {year}
  {2013})}\BibitemShut {NoStop}%
\bibitem [{\citenamefont {Schaibley}\ \emph {et~al.}(2016)\citenamefont
  {Schaibley}, \citenamefont {Yu}, \citenamefont {Clark}, \citenamefont
  {Rivera}, \citenamefont {Ross}, \citenamefont {Seyler}, \citenamefont {Yao},\
  and\ \citenamefont {Xu}}]{Schaibley2016}%
  \BibitemOpen
  \bibfield  {author} {\bibinfo {author} {\bibfnamefont {J.~R.}\ \bibnamefont
  {Schaibley}}, \bibinfo {author} {\bibfnamefont {H.}~\bibnamefont {Yu}},
  \bibinfo {author} {\bibfnamefont {G.}~\bibnamefont {Clark}}, \bibinfo
  {author} {\bibfnamefont {P.}~\bibnamefont {Rivera}}, \bibinfo {author}
  {\bibfnamefont {J.~S.}\ \bibnamefont {Ross}}, \bibinfo {author}
  {\bibfnamefont {K.~L.}\ \bibnamefont {Seyler}}, \bibinfo {author}
  {\bibfnamefont {W.}~\bibnamefont {Yao}},\ and\ \bibinfo {author}
  {\bibfnamefont {X.}~\bibnamefont {Xu}},\ }\href
  {https://doi.org/10.1038/natrevmats.2016.55} {\bibfield  {journal} {\bibinfo
  {journal} {Nat. Rev. Mater.}\ }\textbf {\bibinfo {volume} {1}},\ \bibinfo
  {pages} {16055} (\bibinfo {year} {2016})}\BibitemShut {NoStop}%
\bibitem [{\citenamefont {Islam}\ and\ \citenamefont
  {Benjamin}(2016)}]{ISLAM2016304}%
  \BibitemOpen
  \bibfield  {author} {\bibinfo {author} {\bibfnamefont {S.~F.}\ \bibnamefont
  {Islam}}\ and\ \bibinfo {author} {\bibfnamefont {C.}~\bibnamefont
  {Benjamin}},\ }\href
  {https://doi.org/https://doi.org/10.1016/j.carbon.2016.09.025} {\bibfield
  {journal} {\bibinfo  {journal} {Carbon}\ }\textbf {\bibinfo {volume} {110}},\
  \bibinfo {pages} {304} (\bibinfo {year} {2016})}\BibitemShut {NoStop}%
\bibitem [{\citenamefont {Xu}\ \emph {et~al.}(2020)\citenamefont {Xu},
  \citenamefont {Zhou}, \citenamefont {Scharf},\ and\ \citenamefont {\ifmmode
  \check{Z}\else \v{Z}\fi{}uti\ifmmode~\acute{c}\else
  \'{c}\fi{}}}]{PhysRevLett.125.157402}%
  \BibitemOpen
  \bibfield  {author} {\bibinfo {author} {\bibfnamefont {G.}~\bibnamefont
  {Xu}}, \bibinfo {author} {\bibfnamefont {T.}~\bibnamefont {Zhou}}, \bibinfo
  {author} {\bibfnamefont {B.}~\bibnamefont {Scharf}},\ and\ \bibinfo {author}
  {\bibfnamefont {I.}~\bibnamefont {\ifmmode \check{Z}\else
  \v{Z}\fi{}uti\ifmmode~\acute{c}\else \'{c}\fi{}}},\ }\href
  {https://doi.org/10.1103/PhysRevLett.125.157402} {\bibfield  {journal}
  {\bibinfo  {journal} {Phys. Rev. Lett.}\ }\textbf {\bibinfo {volume} {125}},\
  \bibinfo {pages} {157402} (\bibinfo {year} {2020})}\BibitemShut {NoStop}%
\bibitem [{\citenamefont {Zhou}\ \emph {et~al.}(2021)\citenamefont {Zhou},
  \citenamefont {Cheng}, \citenamefont {Schleenvoigt}, \citenamefont
  {Sch\"uffelgen}, \citenamefont {Jiang}, \citenamefont {Yang},\ and\
  \citenamefont {\ifmmode \check{Z}\else \v{Z}\fi{}uti\ifmmode~\acute{c}\else
  \'{c}\fi{}}}]{PhysRevLett.127.116402}%
  \BibitemOpen
  \bibfield  {author} {\bibinfo {author} {\bibfnamefont {T.}~\bibnamefont
  {Zhou}}, \bibinfo {author} {\bibfnamefont {S.}~\bibnamefont {Cheng}},
  \bibinfo {author} {\bibfnamefont {M.}~\bibnamefont {Schleenvoigt}}, \bibinfo
  {author} {\bibfnamefont {P.}~\bibnamefont {Sch\"uffelgen}}, \bibinfo {author}
  {\bibfnamefont {H.}~\bibnamefont {Jiang}}, \bibinfo {author} {\bibfnamefont
  {Z.}~\bibnamefont {Yang}},\ and\ \bibinfo {author} {\bibfnamefont
  {I.}~\bibnamefont {\ifmmode \check{Z}\else
  \v{Z}\fi{}uti\ifmmode~\acute{c}\else \'{c}\fi{}}},\ }\href
  {https://doi.org/10.1103/PhysRevLett.127.116402} {\bibfield  {journal}
  {\bibinfo  {journal} {Phys. Rev. Lett.}\ }\textbf {\bibinfo {volume} {127}},\
  \bibinfo {pages} {116402} (\bibinfo {year} {2021})}\BibitemShut {NoStop}%
\bibitem [{\citenamefont {Zhang}\ \emph {et~al.}(2011)\citenamefont {Zhang},
  \citenamefont {Jung}, \citenamefont {Fiete}, \citenamefont {Niu},\ and\
  \citenamefont {MacDonald}}]{PhysRevLett.106.156801}%
  \BibitemOpen
  \bibfield  {author} {\bibinfo {author} {\bibfnamefont {F.}~\bibnamefont
  {Zhang}}, \bibinfo {author} {\bibfnamefont {J.}~\bibnamefont {Jung}},
  \bibinfo {author} {\bibfnamefont {G.~A.}\ \bibnamefont {Fiete}}, \bibinfo
  {author} {\bibfnamefont {Q.}~\bibnamefont {Niu}},\ and\ \bibinfo {author}
  {\bibfnamefont {A.~H.}\ \bibnamefont {MacDonald}},\ }\href
  {https://doi.org/10.1103/PhysRevLett.106.156801} {\bibfield  {journal}
  {\bibinfo  {journal} {Phys. Rev. Lett.}\ }\textbf {\bibinfo {volume} {106}},\
  \bibinfo {pages} {156801} (\bibinfo {year} {2011})}\BibitemShut {NoStop}%
\bibitem [{\citenamefont {Zhang}\ \emph {et~al.}(2013)\citenamefont {Zhang},
  \citenamefont {MacDonald},\ and\ \citenamefont {Mele}}]{Zhang10546}%
  \BibitemOpen
  \bibfield  {author} {\bibinfo {author} {\bibfnamefont {F.}~\bibnamefont
  {Zhang}}, \bibinfo {author} {\bibfnamefont {A.~H.}\ \bibnamefont
  {MacDonald}},\ and\ \bibinfo {author} {\bibfnamefont {E.~J.}\ \bibnamefont
  {Mele}},\ }\href {https://doi.org/10.1073/pnas.1308853110} {\bibfield
  {journal} {\bibinfo  {journal} {Proc. Natl. Acad. Sci. USA}\ }\textbf
  {\bibinfo {volume} {110}},\ \bibinfo {pages} {10546} (\bibinfo {year}
  {2013})}\BibitemShut {NoStop}%
\bibitem [{\citenamefont {Ezawa}(2013{\natexlab{a}})}]{PhysRevB.88.161406}%
  \BibitemOpen
  \bibfield  {author} {\bibinfo {author} {\bibfnamefont {M.}~\bibnamefont
  {Ezawa}},\ }\href {https://doi.org/10.1103/PhysRevB.88.161406} {\bibfield
  {journal} {\bibinfo  {journal} {Phys. Rev. B}\ }\textbf {\bibinfo {volume}
  {88}},\ \bibinfo {pages} {161406} (\bibinfo {year}
  {2013}{\natexlab{a}})}\BibitemShut {NoStop}%
\bibitem [{\citenamefont {Pan}\ \emph {et~al.}(2014)\citenamefont {Pan},
  \citenamefont {Li}, \citenamefont {Liu}, \citenamefont {Zhu}, \citenamefont
  {Qiao},\ and\ \citenamefont {Yao}}]{PhysRevLett.112.106802}%
  \BibitemOpen
  \bibfield  {author} {\bibinfo {author} {\bibfnamefont {H.}~\bibnamefont
  {Pan}}, \bibinfo {author} {\bibfnamefont {Z.}~\bibnamefont {Li}}, \bibinfo
  {author} {\bibfnamefont {C.-C.}\ \bibnamefont {Liu}}, \bibinfo {author}
  {\bibfnamefont {G.}~\bibnamefont {Zhu}}, \bibinfo {author} {\bibfnamefont
  {Z.}~\bibnamefont {Qiao}},\ and\ \bibinfo {author} {\bibfnamefont
  {Y.}~\bibnamefont {Yao}},\ }\href
  {https://doi.org/10.1103/PhysRevLett.112.106802} {\bibfield  {journal}
  {\bibinfo  {journal} {Phys. Rev. Lett.}\ }\textbf {\bibinfo {volume} {112}},\
  \bibinfo {pages} {106802} (\bibinfo {year} {2014})}\BibitemShut {NoStop}%
\bibitem [{\citenamefont {Zhou}\ \emph {et~al.}(2017)\citenamefont {Zhou},
  \citenamefont {Sun},\ and\ \citenamefont {Jena}}]{PhysRevLett.119.046403}%
  \BibitemOpen
  \bibfield  {author} {\bibinfo {author} {\bibfnamefont {J.}~\bibnamefont
  {Zhou}}, \bibinfo {author} {\bibfnamefont {Q.}~\bibnamefont {Sun}},\ and\
  \bibinfo {author} {\bibfnamefont {P.}~\bibnamefont {Jena}},\ }\href
  {https://doi.org/10.1103/PhysRevLett.119.046403} {\bibfield  {journal}
  {\bibinfo  {journal} {Phys. Rev. Lett.}\ }\textbf {\bibinfo {volume} {119}},\
  \bibinfo {pages} {046403} (\bibinfo {year} {2017})}\BibitemShut {NoStop}%
\bibitem [{\citenamefont {Li}\ \emph {et~al.}(2018)\citenamefont {Li},
  \citenamefont {Liu}, \citenamefont {Wang}, \citenamefont {Wang},
  \citenamefont {Xu},\ and\ \citenamefont {Duan}}]{PhysRevB.98.201407}%
  \BibitemOpen
  \bibfield  {author} {\bibinfo {author} {\bibfnamefont {Y.}~\bibnamefont
  {Li}}, \bibinfo {author} {\bibfnamefont {Y.}~\bibnamefont {Liu}}, \bibinfo
  {author} {\bibfnamefont {C.}~\bibnamefont {Wang}}, \bibinfo {author}
  {\bibfnamefont {J.}~\bibnamefont {Wang}}, \bibinfo {author} {\bibfnamefont
  {Y.}~\bibnamefont {Xu}},\ and\ \bibinfo {author} {\bibfnamefont
  {W.}~\bibnamefont {Duan}},\ }\href
  {https://doi.org/10.1103/PhysRevB.98.201407} {\bibfield  {journal} {\bibinfo
  {journal} {Phys. Rev. B}\ }\textbf {\bibinfo {volume} {98}},\ \bibinfo
  {pages} {201407} (\bibinfo {year} {2018})}\BibitemShut {NoStop}%
\bibitem [{\citenamefont {Pan}\ \emph {et~al.}(2015)\citenamefont {Pan},
  \citenamefont {Li}, \citenamefont {Jiang}, \citenamefont {Yao},\ and\
  \citenamefont {Yang}}]{PhysRevB.91.045404}%
  \BibitemOpen
  \bibfield  {author} {\bibinfo {author} {\bibfnamefont {H.}~\bibnamefont
  {Pan}}, \bibinfo {author} {\bibfnamefont {X.}~\bibnamefont {Li}}, \bibinfo
  {author} {\bibfnamefont {H.}~\bibnamefont {Jiang}}, \bibinfo {author}
  {\bibfnamefont {Y.}~\bibnamefont {Yao}},\ and\ \bibinfo {author}
  {\bibfnamefont {S.~A.}\ \bibnamefont {Yang}},\ }\href
  {https://doi.org/10.1103/PhysRevB.91.045404} {\bibfield  {journal} {\bibinfo
  {journal} {Phys. Rev. B}\ }\textbf {\bibinfo {volume} {91}},\ \bibinfo
  {pages} {045404} (\bibinfo {year} {2015})}\BibitemShut {NoStop}%
\bibitem [{\citenamefont {Zhai}\ and\ \citenamefont
  {Blanter}(2020)}]{PhysRevB.101.155425}%
  \BibitemOpen
  \bibfield  {author} {\bibinfo {author} {\bibfnamefont {X.}~\bibnamefont
  {Zhai}}\ and\ \bibinfo {author} {\bibfnamefont {Y.~M.}\ \bibnamefont
  {Blanter}},\ }\href {https://doi.org/10.1103/PhysRevB.101.155425} {\bibfield
  {journal} {\bibinfo  {journal} {Phys. Rev. B}\ }\textbf {\bibinfo {volume}
  {101}},\ \bibinfo {pages} {155425} (\bibinfo {year} {2020})}\BibitemShut
  {NoStop}%
\bibitem [{\citenamefont {Vila}\ \emph {et~al.}(2021)\citenamefont {Vila},
  \citenamefont {Garcia},\ and\ \citenamefont {Roche}}]{PhysRevB.104.L161113}%
  \BibitemOpen
  \bibfield  {author} {\bibinfo {author} {\bibfnamefont {M.}~\bibnamefont
  {Vila}}, \bibinfo {author} {\bibfnamefont {J.~H.}\ \bibnamefont {Garcia}},\
  and\ \bibinfo {author} {\bibfnamefont {S.}~\bibnamefont {Roche}},\ }\href
  {https://doi.org/10.1103/PhysRevB.104.L161113} {\bibfield  {journal}
  {\bibinfo  {journal} {Phys. Rev. B}\ }\textbf {\bibinfo {volume} {104}},\
  \bibinfo {pages} {L161113} (\bibinfo {year} {2021})}\BibitemShut {NoStop}%
\bibitem [{\citenamefont {Zhou}\ \emph {et~al.}(2015)\citenamefont {Zhou},
  \citenamefont {Zhang}, \citenamefont {Zhao}, \citenamefont {Zhang},\ and\
  \citenamefont {Yang}}]{nanolett.5b01373}%
  \BibitemOpen
  \bibfield  {author} {\bibinfo {author} {\bibfnamefont {T.}~\bibnamefont
  {Zhou}}, \bibinfo {author} {\bibfnamefont {J.}~\bibnamefont {Zhang}},
  \bibinfo {author} {\bibfnamefont {B.}~\bibnamefont {Zhao}}, \bibinfo {author}
  {\bibfnamefont {H.}~\bibnamefont {Zhang}},\ and\ \bibinfo {author}
  {\bibfnamefont {Z.}~\bibnamefont {Yang}},\ }\href
  {https://doi.org/10.1021/acs.nanolett.5b01373} {\bibfield  {journal}
  {\bibinfo  {journal} {Nano Lett.}\ }\textbf {\bibinfo {volume} {15}},\
  \bibinfo {pages} {5149} (\bibinfo {year} {2015})}\BibitemShut {NoStop}%
\bibitem [{\citenamefont {Liu}\ \emph {et~al.}(2015)\citenamefont {Liu},
  \citenamefont {Zhou},\ and\ \citenamefont {Yao}}]{PhysRevB.91.165430}%
  \BibitemOpen
  \bibfield  {author} {\bibinfo {author} {\bibfnamefont {C.-C.}\ \bibnamefont
  {Liu}}, \bibinfo {author} {\bibfnamefont {J.-J.}\ \bibnamefont {Zhou}},\ and\
  \bibinfo {author} {\bibfnamefont {Y.}~\bibnamefont {Yao}},\ }\href
  {https://doi.org/10.1103/PhysRevB.91.165430} {\bibfield  {journal} {\bibinfo
  {journal} {Phys. Rev. B}\ }\textbf {\bibinfo {volume} {91}},\ \bibinfo
  {pages} {165430} (\bibinfo {year} {2015})}\BibitemShut {NoStop}%
\bibitem [{\citenamefont {Niu}\ \emph {et~al.}(2015)\citenamefont {Niu},
  \citenamefont {Bihlmayer}, \citenamefont {Zhang}, \citenamefont {Wortmann},
  \citenamefont {Bl\"ugel},\ and\ \citenamefont
  {Mokrousov}}]{PhysRevB.91.041303}%
  \BibitemOpen
  \bibfield  {author} {\bibinfo {author} {\bibfnamefont {C.}~\bibnamefont
  {Niu}}, \bibinfo {author} {\bibfnamefont {G.}~\bibnamefont {Bihlmayer}},
  \bibinfo {author} {\bibfnamefont {H.}~\bibnamefont {Zhang}}, \bibinfo
  {author} {\bibfnamefont {D.}~\bibnamefont {Wortmann}}, \bibinfo {author}
  {\bibfnamefont {S.}~\bibnamefont {Bl\"ugel}},\ and\ \bibinfo {author}
  {\bibfnamefont {Y.}~\bibnamefont {Mokrousov}},\ }\href
  {https://doi.org/10.1103/PhysRevB.91.041303} {\bibfield  {journal} {\bibinfo
  {journal} {Phys. Rev. B}\ }\textbf {\bibinfo {volume} {91}},\ \bibinfo
  {pages} {041303} (\bibinfo {year} {2015})}\BibitemShut {NoStop}%
\bibitem [{\citenamefont {Zhou}\ \emph {et~al.}(2016)\citenamefont {Zhou},
  \citenamefont {Zhang}, \citenamefont {Xue}, \citenamefont {Zhao},
  \citenamefont {Zhang}, \citenamefont {Jiang},\ and\ \citenamefont
  {Yang}}]{PhysRevB.94.235449}%
  \BibitemOpen
  \bibfield  {author} {\bibinfo {author} {\bibfnamefont {T.}~\bibnamefont
  {Zhou}}, \bibinfo {author} {\bibfnamefont {J.}~\bibnamefont {Zhang}},
  \bibinfo {author} {\bibfnamefont {Y.}~\bibnamefont {Xue}}, \bibinfo {author}
  {\bibfnamefont {B.}~\bibnamefont {Zhao}}, \bibinfo {author} {\bibfnamefont
  {H.}~\bibnamefont {Zhang}}, \bibinfo {author} {\bibfnamefont
  {H.}~\bibnamefont {Jiang}},\ and\ \bibinfo {author} {\bibfnamefont
  {Z.}~\bibnamefont {Yang}},\ }\href
  {https://doi.org/10.1103/PhysRevB.94.235449} {\bibfield  {journal} {\bibinfo
  {journal} {Phys. Rev. B}\ }\textbf {\bibinfo {volume} {94}},\ \bibinfo
  {pages} {235449} (\bibinfo {year} {2016})}\BibitemShut {NoStop}%
\bibitem [{\citenamefont {Sui}\ \emph {et~al.}(2020)\citenamefont {Sui},
  \citenamefont {Zhang}, \citenamefont {Jin}, \citenamefont {Xia},\ and\
  \citenamefont {Li}}]{QianSui97301}%
  \BibitemOpen
  \bibfield  {author} {\bibinfo {author} {\bibfnamefont {Q.}~\bibnamefont
  {Sui}}, \bibinfo {author} {\bibfnamefont {J.}~\bibnamefont {Zhang}}, \bibinfo
  {author} {\bibfnamefont {S.}~\bibnamefont {Jin}}, \bibinfo {author}
  {\bibfnamefont {Y.}~\bibnamefont {Xia}},\ and\ \bibinfo {author}
  {\bibfnamefont {G.}~\bibnamefont {Li}},\ }\href
  {https://doi.org/10.1088/0256-307X/37/9/097301} {\bibfield  {journal}
  {\bibinfo  {journal} {Chin. Phys. Lett.}\ }\textbf {\bibinfo {volume} {37}},\
  \bibinfo {eid} {097301} (\bibinfo {year} {2020})}\BibitemShut {NoStop}%
\bibitem [{\citenamefont {Hu}\ \emph {et~al.}(2020)\citenamefont {Hu},
  \citenamefont {Tong}, \citenamefont {Shen}, \citenamefont {Wan},\ and\
  \citenamefont {Duan}}]{WOSConcepts}%
  \BibitemOpen
  \bibfield  {author} {\bibinfo {author} {\bibfnamefont {H.}~\bibnamefont
  {Hu}}, \bibinfo {author} {\bibfnamefont {W.-Y.}\ \bibnamefont {Tong}},
  \bibinfo {author} {\bibfnamefont {Y.-H.}\ \bibnamefont {Shen}}, \bibinfo
  {author} {\bibfnamefont {X.}~\bibnamefont {Wan}},\ and\ \bibinfo {author}
  {\bibfnamefont {C.-G.}\ \bibnamefont {Duan}},\ }\href
  {https://doi.org/10.1038/s41524-020-00397-1} {\bibfield  {journal} {\bibinfo
  {journal} {npj Comput. Mater.}\ }\textbf {\bibinfo {volume} {6}},\ \bibinfo
  {pages} {129} (\bibinfo {year} {2020})}\BibitemShut {NoStop}%
\bibitem [{\citenamefont {Liu}\ \emph {et~al.}(2021)\citenamefont {Liu},
  \citenamefont {Han}, \citenamefont {Ren}, \citenamefont {Niu},\ and\
  \citenamefont {Qiao}}]{PhysRevB.104.L121403}%
  \BibitemOpen
  \bibfield  {author} {\bibinfo {author} {\bibfnamefont {Z.}~\bibnamefont
  {Liu}}, \bibinfo {author} {\bibfnamefont {Y.}~\bibnamefont {Han}}, \bibinfo
  {author} {\bibfnamefont {Y.}~\bibnamefont {Ren}}, \bibinfo {author}
  {\bibfnamefont {Q.}~\bibnamefont {Niu}},\ and\ \bibinfo {author}
  {\bibfnamefont {Z.}~\bibnamefont {Qiao}},\ }\href
  {https://doi.org/10.1103/PhysRevB.104.L121403} {\bibfield  {journal}
  {\bibinfo  {journal} {Phys. Rev. B}\ }\textbf {\bibinfo {volume} {104}},\
  \bibinfo {pages} {L121403} (\bibinfo {year} {2021})}\BibitemShut {NoStop}%
\bibitem [{\citenamefont {Mermin}\ and\ \citenamefont
  {Wagner}(1966)}]{PhysRevLett.17.1133}%
  \BibitemOpen
  \bibfield  {author} {\bibinfo {author} {\bibfnamefont {N.~D.}\ \bibnamefont
  {Mermin}}\ and\ \bibinfo {author} {\bibfnamefont {H.}~\bibnamefont
  {Wagner}},\ }\href {https://doi.org/10.1103/PhysRevLett.17.1133} {\bibfield
  {journal} {\bibinfo  {journal} {Phys. Rev. Lett.}\ }\textbf {\bibinfo
  {volume} {17}},\ \bibinfo {pages} {1133} (\bibinfo {year}
  {1966})}\BibitemShut {NoStop}%
\bibitem [{\citenamefont {Wang}\ \emph {et~al.}(2013)\citenamefont {Wang},
  \citenamefont {Steinberg}, \citenamefont {Jarillo-Herrero},\ and\
  \citenamefont {Gedik}}]{science.1239834}%
  \BibitemOpen
  \bibfield  {author} {\bibinfo {author} {\bibfnamefont {Y.~H.}\ \bibnamefont
  {Wang}}, \bibinfo {author} {\bibfnamefont {H.}~\bibnamefont {Steinberg}},
  \bibinfo {author} {\bibfnamefont {P.}~\bibnamefont {Jarillo-Herrero}},\ and\
  \bibinfo {author} {\bibfnamefont {N.}~\bibnamefont {Gedik}},\ }\href
  {https://doi.org/10.1126/science.1239834} {\bibfield  {journal} {\bibinfo
  {journal} {Science}\ }\textbf {\bibinfo {volume} {342}},\ \bibinfo {pages}
  {453} (\bibinfo {year} {2013})}\BibitemShut {NoStop}%
\bibitem [{\citenamefont {Xu}\ \emph {et~al.}(2021)\citenamefont {Xu},
  \citenamefont {Zhou},\ and\ \citenamefont {Li}}]{advs.202101508}%
  \BibitemOpen
  \bibfield  {author} {\bibinfo {author} {\bibfnamefont {H.}~\bibnamefont
  {Xu}}, \bibinfo {author} {\bibfnamefont {J.}~\bibnamefont {Zhou}},\ and\
  \bibinfo {author} {\bibfnamefont {J.}~\bibnamefont {Li}},\ }\href
  {https://doi.org/10.1002/advs.202101508} {\bibfield  {journal} {\bibinfo
  {journal} {Adv. Sci.}\ }\textbf {\bibinfo {volume} {8}},\ \bibinfo {pages}
  {2101508} (\bibinfo {year} {2021})}\BibitemShut {NoStop}%
\bibitem [{\citenamefont {Kundu}\ \emph {et~al.}(2016)\citenamefont {Kundu},
  \citenamefont {Fertig},\ and\ \citenamefont
  {Seradjeh}}]{PhysRevLett.116.016802}%
  \BibitemOpen
  \bibfield  {author} {\bibinfo {author} {\bibfnamefont {A.}~\bibnamefont
  {Kundu}}, \bibinfo {author} {\bibfnamefont {H.~A.}\ \bibnamefont {Fertig}},\
  and\ \bibinfo {author} {\bibfnamefont {B.}~\bibnamefont {Seradjeh}},\ }\href
  {https://doi.org/10.1103/PhysRevLett.116.016802} {\bibfield  {journal}
  {\bibinfo  {journal} {Phys. Rev. Lett.}\ }\textbf {\bibinfo {volume} {116}},\
  \bibinfo {pages} {016802} (\bibinfo {year} {2016})}\BibitemShut {NoStop}%
\bibitem [{\citenamefont {Lindner}\ \emph {et~al.}(2011)\citenamefont
  {Lindner}, \citenamefont {Refael},\ and\ \citenamefont
  {Galitski}}]{lindner2011floquet}%
  \BibitemOpen
  \bibfield  {author} {\bibinfo {author} {\bibfnamefont {N.~H.}\ \bibnamefont
  {Lindner}}, \bibinfo {author} {\bibfnamefont {G.}~\bibnamefont {Refael}},\
  and\ \bibinfo {author} {\bibfnamefont {V.}~\bibnamefont {Galitski}},\ }\href
  {https://doi.org/10.1038/nphys1926} {\bibfield  {journal} {\bibinfo
  {journal} {Nat. Phys.}\ }\textbf {\bibinfo {volume} {7}},\ \bibinfo {pages}
  {490} (\bibinfo {year} {2011})}\BibitemShut {NoStop}%
\bibitem [{\citenamefont {Usaj}\ \emph {et~al.}(2014)\citenamefont {Usaj},
  \citenamefont {Perez-Piskunow}, \citenamefont {Foa~Torres},\ and\
  \citenamefont {Balseiro}}]{PhysRevB.90.115423}%
  \BibitemOpen
  \bibfield  {author} {\bibinfo {author} {\bibfnamefont {G.}~\bibnamefont
  {Usaj}}, \bibinfo {author} {\bibfnamefont {P.~M.}\ \bibnamefont
  {Perez-Piskunow}}, \bibinfo {author} {\bibfnamefont {L.~E.~F.}\ \bibnamefont
  {Foa~Torres}},\ and\ \bibinfo {author} {\bibfnamefont {C.~A.}\ \bibnamefont
  {Balseiro}},\ }\href {https://doi.org/10.1103/PhysRevB.90.115423} {\bibfield
  {journal} {\bibinfo  {journal} {Phys. Rev. B}\ }\textbf {\bibinfo {volume}
  {90}},\ \bibinfo {pages} {115423} (\bibinfo {year} {2014})}\BibitemShut
  {NoStop}%
\bibitem [{\citenamefont {Ezawa}(2013{\natexlab{b}})}]{PhysRevLett.110.026603}%
  \BibitemOpen
  \bibfield  {author} {\bibinfo {author} {\bibfnamefont {M.}~\bibnamefont
  {Ezawa}},\ }\href {https://doi.org/10.1103/PhysRevLett.110.026603} {\bibfield
   {journal} {\bibinfo  {journal} {Phys. Rev. Lett.}\ }\textbf {\bibinfo
  {volume} {110}},\ \bibinfo {pages} {026603} (\bibinfo {year}
  {2013}{\natexlab{b}})}\BibitemShut {NoStop}%
\bibitem [{\citenamefont {G\'omez-Le\'on}\ and\ \citenamefont
  {Platero}(2013)}]{PhysRevLett.110.200403}%
  \BibitemOpen
  \bibfield  {author} {\bibinfo {author} {\bibfnamefont {A.}~\bibnamefont
  {G\'omez-Le\'on}}\ and\ \bibinfo {author} {\bibfnamefont {G.}~\bibnamefont
  {Platero}},\ }\href {https://doi.org/10.1103/PhysRevLett.110.200403}
  {\bibfield  {journal} {\bibinfo  {journal} {Phys. Rev. Lett.}\ }\textbf
  {\bibinfo {volume} {110}},\ \bibinfo {pages} {200403} (\bibinfo {year}
  {2013})}\BibitemShut {NoStop}%
\bibitem [{\citenamefont {Yan}\ and\ \citenamefont
  {Wang}(2016)}]{PhysRevLett.117.087402}%
  \BibitemOpen
  \bibfield  {author} {\bibinfo {author} {\bibfnamefont {Z.}~\bibnamefont
  {Yan}}\ and\ \bibinfo {author} {\bibfnamefont {Z.}~\bibnamefont {Wang}},\
  }\href {https://doi.org/10.1103/PhysRevLett.117.087402} {\bibfield  {journal}
  {\bibinfo  {journal} {Phys. Rev. Lett.}\ }\textbf {\bibinfo {volume} {117}},\
  \bibinfo {pages} {087402} (\bibinfo {year} {2016})}\BibitemShut {NoStop}%
\bibitem [{\citenamefont {Chen}\ \emph {et~al.}(2018)\citenamefont {Chen},
  \citenamefont {Zhou},\ and\ \citenamefont {Xu}}]{PhysRevB.97.155152}%
  \BibitemOpen
  \bibfield  {author} {\bibinfo {author} {\bibfnamefont {R.}~\bibnamefont
  {Chen}}, \bibinfo {author} {\bibfnamefont {B.}~\bibnamefont {Zhou}},\ and\
  \bibinfo {author} {\bibfnamefont {D.-H.}\ \bibnamefont {Xu}},\ }\href
  {https://doi.org/10.1103/PhysRevB.97.155152} {\bibfield  {journal} {\bibinfo
  {journal} {Phys. Rev. B}\ }\textbf {\bibinfo {volume} {97}},\ \bibinfo
  {pages} {155152} (\bibinfo {year} {2018})}\BibitemShut {NoStop}%
\bibitem [{\citenamefont {Liu}\ \emph {et~al.}(2018)\citenamefont {Liu},
  \citenamefont {Sun}, \citenamefont {Cheng}, \citenamefont {Liu},\ and\
  \citenamefont {Meng}}]{PhysRevLett.120.237403}%
  \BibitemOpen
  \bibfield  {author} {\bibinfo {author} {\bibfnamefont {H.}~\bibnamefont
  {Liu}}, \bibinfo {author} {\bibfnamefont {J.-T.}\ \bibnamefont {Sun}},
  \bibinfo {author} {\bibfnamefont {C.}~\bibnamefont {Cheng}}, \bibinfo
  {author} {\bibfnamefont {F.}~\bibnamefont {Liu}},\ and\ \bibinfo {author}
  {\bibfnamefont {S.}~\bibnamefont {Meng}},\ }\href
  {https://doi.org/10.1103/PhysRevLett.120.237403} {\bibfield  {journal}
  {\bibinfo  {journal} {Phys. Rev. Lett.}\ }\textbf {\bibinfo {volume} {120}},\
  \bibinfo {pages} {237403} (\bibinfo {year} {2018})}\BibitemShut {NoStop}%
\bibitem [{\citenamefont {H{\"u}bener}\ \emph {et~al.}(2017)\citenamefont
  {H{\"u}bener}, \citenamefont {Sentef}, \citenamefont {De~Giovannini},
  \citenamefont {Kemper},\ and\ \citenamefont {Rubio}}]{hubener2017creating}%
  \BibitemOpen
  \bibfield  {author} {\bibinfo {author} {\bibfnamefont {H.}~\bibnamefont
  {H{\"u}bener}}, \bibinfo {author} {\bibfnamefont {M.~A.}\ \bibnamefont
  {Sentef}}, \bibinfo {author} {\bibfnamefont {U.}~\bibnamefont
  {De~Giovannini}}, \bibinfo {author} {\bibfnamefont {A.~F.}\ \bibnamefont
  {Kemper}},\ and\ \bibinfo {author} {\bibfnamefont {A.}~\bibnamefont
  {Rubio}},\ }\href {https://doi.org/10.1038/ncomms13940} {\bibfield  {journal}
  {\bibinfo  {journal} {Nat. Commun.}\ }\textbf {\bibinfo {volume} {8}},\
  \bibinfo {pages} {1} (\bibinfo {year} {2017})}\BibitemShut {NoStop}%
\bibitem [{\citenamefont {McIver}\ \emph {et~al.}(2020)\citenamefont {McIver},
  \citenamefont {Schulte}, \citenamefont {Stein}, \citenamefont {Matsuyama},
  \citenamefont {Jotzu}, \citenamefont {Meier},\ and\ \citenamefont
  {Cavalleri}}]{mciver2020light}%
  \BibitemOpen
  \bibfield  {author} {\bibinfo {author} {\bibfnamefont {J.~W.}\ \bibnamefont
  {McIver}}, \bibinfo {author} {\bibfnamefont {B.}~\bibnamefont {Schulte}},
  \bibinfo {author} {\bibfnamefont {F.-U.}\ \bibnamefont {Stein}}, \bibinfo
  {author} {\bibfnamefont {T.}~\bibnamefont {Matsuyama}}, \bibinfo {author}
  {\bibfnamefont {G.}~\bibnamefont {Jotzu}}, \bibinfo {author} {\bibfnamefont
  {G.}~\bibnamefont {Meier}},\ and\ \bibinfo {author} {\bibfnamefont
  {A.}~\bibnamefont {Cavalleri}},\ }\href
  {https://doi.org/10.1038/s41567-019-0698-y} {\bibfield  {journal} {\bibinfo
  {journal} {Nat. Phys.}\ }\textbf {\bibinfo {volume} {16}},\ \bibinfo {pages}
  {38} (\bibinfo {year} {2020})}\BibitemShut {NoStop}%
\bibitem [{\citenamefont {Cayssol}\ \emph {et~al.}(2013)\citenamefont
  {Cayssol}, \citenamefont {D{\'o}ra}, \citenamefont {Simon},\ and\
  \citenamefont {Moessner}}]{pssr.201206451}%
  \BibitemOpen
  \bibfield  {author} {\bibinfo {author} {\bibfnamefont {J.}~\bibnamefont
  {Cayssol}}, \bibinfo {author} {\bibfnamefont {B.}~\bibnamefont {D{\'o}ra}},
  \bibinfo {author} {\bibfnamefont {F.}~\bibnamefont {Simon}},\ and\ \bibinfo
  {author} {\bibfnamefont {R.}~\bibnamefont {Moessner}},\ }\href
  {https://doi.org/https://doi.org/10.1002/pssr.201206451} {\bibfield
  {journal} {\bibinfo  {journal} {Phys. Status Solidi RRL}\ }\textbf {\bibinfo
  {volume} {7}},\ \bibinfo {pages} {101} (\bibinfo {year} {2013})}\BibitemShut
  {NoStop}%
\bibitem [{\citenamefont {Oka}\ and\ \citenamefont
  {Kitamura}(2019)}]{annurev-conmatphys-031218-013423}%
  \BibitemOpen
  \bibfield  {author} {\bibinfo {author} {\bibfnamefont {T.}~\bibnamefont
  {Oka}}\ and\ \bibinfo {author} {\bibfnamefont {S.}~\bibnamefont {Kitamura}},\
  }\href {https://doi.org/10.1146/annurev-conmatphys-031218-013423} {\bibfield
  {journal} {\bibinfo  {journal} {Annu. Rev. Condens. Matter Phys.}\ }\textbf
  {\bibinfo {volume} {10}},\ \bibinfo {pages} {387} (\bibinfo {year}
  {2019})}\BibitemShut {NoStop}%
\bibitem [{\citenamefont {Tong}\ \emph {et~al.}(2017)\citenamefont {Tong},
  \citenamefont {Yu}, \citenamefont {Zhu}, \citenamefont {Wang}, \citenamefont
  {Xu},\ and\ \citenamefont {Yao}}]{2017Topological}%
  \BibitemOpen
  \bibfield  {author} {\bibinfo {author} {\bibfnamefont {Q.}~\bibnamefont
  {Tong}}, \bibinfo {author} {\bibfnamefont {H.}~\bibnamefont {Yu}}, \bibinfo
  {author} {\bibfnamefont {Q.}~\bibnamefont {Zhu}}, \bibinfo {author}
  {\bibfnamefont {Y.}~\bibnamefont {Wang}}, \bibinfo {author} {\bibfnamefont
  {X.}~\bibnamefont {Xu}},\ and\ \bibinfo {author} {\bibfnamefont
  {W.}~\bibnamefont {Yao}},\ }\href {https://doi.org/10.1038/nphys3968}
  {\bibfield  {journal} {\bibinfo  {journal} {Nat. Phys.}\ }\textbf {\bibinfo
  {volume} {13}},\ \bibinfo {pages} {356-362} (\bibinfo {year}
  {2017})}\BibitemShut {NoStop}%
\bibitem [{SM()}]{SM}%
  \BibitemOpen
  \href@noop {} {}\bibinfo {note} {Supplemental Material}\BibitemShut {NoStop}%
\bibitem [{\citenamefont {Milfeld}\ and\ \citenamefont
  {Wyatt}(1983)}]{PhysRevA.27.72}%
  \BibitemOpen
  \bibfield  {author} {\bibinfo {author} {\bibfnamefont {K.~F.}\ \bibnamefont
  {Milfeld}}\ and\ \bibinfo {author} {\bibfnamefont {R.~E.}\ \bibnamefont
  {Wyatt}},\ }\href {https://doi.org/10.1103/PhysRevA.27.72} {\bibfield
  {journal} {\bibinfo  {journal} {Phys. Rev. A}\ }\textbf {\bibinfo {volume}
  {27}},\ \bibinfo {pages} {72} (\bibinfo {year} {1983})}\BibitemShut {NoStop}%
\bibitem [{\citenamefont {Goldman}\ and\ \citenamefont
  {Dalibard}(2014)}]{PhysRevX.4.031027}%
  \BibitemOpen
  \bibfield  {author} {\bibinfo {author} {\bibfnamefont {N.}~\bibnamefont
  {Goldman}}\ and\ \bibinfo {author} {\bibfnamefont {J.}~\bibnamefont
  {Dalibard}},\ }\href {https://doi.org/10.1103/PhysRevX.4.031027} {\bibfield
  {journal} {\bibinfo  {journal} {Phys. Rev. X}\ }\textbf {\bibinfo {volume}
  {4}},\ \bibinfo {pages} {031027} (\bibinfo {year} {2014})}\BibitemShut
  {NoStop}%
\bibitem [{\citenamefont {Yang}\ \emph {et~al.}(2011)\citenamefont {Yang},
  \citenamefont {Xu}, \citenamefont {Sheng}, \citenamefont {Wang},
  \citenamefont {Xing},\ and\ \citenamefont {Sheng}}]{PhysRevLett.107.066602}%
  \BibitemOpen
  \bibfield  {author} {\bibinfo {author} {\bibfnamefont {Y.}~\bibnamefont
  {Yang}}, \bibinfo {author} {\bibfnamefont {Z.}~\bibnamefont {Xu}}, \bibinfo
  {author} {\bibfnamefont {L.}~\bibnamefont {Sheng}}, \bibinfo {author}
  {\bibfnamefont {B.}~\bibnamefont {Wang}}, \bibinfo {author} {\bibfnamefont
  {D.~Y.}\ \bibnamefont {Xing}},\ and\ \bibinfo {author} {\bibfnamefont
  {D.~N.}\ \bibnamefont {Sheng}},\ }\href
  {https://doi.org/10.1103/PhysRevLett.107.066602} {\bibfield  {journal}
  {\bibinfo  {journal} {Phys. Rev. Lett.}\ }\textbf {\bibinfo {volume} {107}},\
  \bibinfo {pages} {066602} (\bibinfo {year} {2011})}\BibitemShut {NoStop}%
\bibitem [{\citenamefont {Kohn}\ and\ \citenamefont
  {Sham}(1965)}]{PhysRev.140.A1133}%
  \BibitemOpen
  \bibfield  {author} {\bibinfo {author} {\bibfnamefont {W.}~\bibnamefont
  {Kohn}}\ and\ \bibinfo {author} {\bibfnamefont {L.~J.}\ \bibnamefont
  {Sham}},\ }\href {https://doi.org/10.1103/PhysRev.140.A1133} {\bibfield
  {journal} {\bibinfo  {journal} {Phys. Rev.}\ }\textbf {\bibinfo {volume}
  {140}},\ \bibinfo {pages} {A1133} (\bibinfo {year} {1965})}\BibitemShut
  {NoStop}%
\bibitem [{\citenamefont {Kresse}\ and\ \citenamefont
  {Furthm\"uller}(1996)}]{PhysRevB.54.11169}%
  \BibitemOpen
  \bibfield  {author} {\bibinfo {author} {\bibfnamefont {G.}~\bibnamefont
  {Kresse}}\ and\ \bibinfo {author} {\bibfnamefont {J.}~\bibnamefont
  {Furthm\"uller}},\ }\href {https://doi.org/10.1103/PhysRevB.54.11169}
  {\bibfield  {journal} {\bibinfo  {journal} {Phys. Rev. B}\ }\textbf {\bibinfo
  {volume} {54}},\ \bibinfo {pages} {11169} (\bibinfo {year}
  {1996})}\BibitemShut {NoStop}%
\bibitem [{\citenamefont {Mostofi}\ \emph {et~al.}(2014)\citenamefont
  {Mostofi}, \citenamefont {Yates}, \citenamefont {Pizzi}, \citenamefont {Lee},
  \citenamefont {Souza}, \citenamefont {Vanderbilt},\ and\ \citenamefont
  {Marzari}}]{Mostofi2014}%
  \BibitemOpen
  \bibfield  {author} {\bibinfo {author} {\bibfnamefont {A.~A.}\ \bibnamefont
  {Mostofi}}, \bibinfo {author} {\bibfnamefont {J.~R.}\ \bibnamefont {Yates}},
  \bibinfo {author} {\bibfnamefont {G.}~\bibnamefont {Pizzi}}, \bibinfo
  {author} {\bibfnamefont {Y.-S.}\ \bibnamefont {Lee}}, \bibinfo {author}
  {\bibfnamefont {I.}~\bibnamefont {Souza}}, \bibinfo {author} {\bibfnamefont
  {D.}~\bibnamefont {Vanderbilt}},\ and\ \bibinfo {author} {\bibfnamefont
  {N.}~\bibnamefont {Marzari}},\ }\href
  {https://doi.org/https://doi.org/10.1016/j.cpc.2014.05.003} {\bibfield
  {journal} {\bibinfo  {journal} {Comput. Phys. Commun.}\ }\textbf {\bibinfo
  {volume} {185}},\ \bibinfo {pages} {2309 } (\bibinfo {year}
  {2014})}\BibitemShut {NoStop}%
\bibitem [{\citenamefont {Deng}\ \emph {et~al.}(2020)\citenamefont {Deng},
  \citenamefont {Zheng}, \citenamefont {Zhan}, \citenamefont {Fan},
  \citenamefont {Wu},\ and\ \citenamefont {Wang}}]{PhysRevB.102.201105}%
  \BibitemOpen
  \bibfield  {author} {\bibinfo {author} {\bibfnamefont {T.}~\bibnamefont
  {Deng}}, \bibinfo {author} {\bibfnamefont {B.}~\bibnamefont {Zheng}},
  \bibinfo {author} {\bibfnamefont {F.}~\bibnamefont {Zhan}}, \bibinfo {author}
  {\bibfnamefont {J.}~\bibnamefont {Fan}}, \bibinfo {author} {\bibfnamefont
  {X.}~\bibnamefont {Wu}},\ and\ \bibinfo {author} {\bibfnamefont
  {R.}~\bibnamefont {Wang}},\ }\href
  {https://doi.org/10.1103/PhysRevB.102.201105} {\bibfield  {journal} {\bibinfo
   {journal} {Phys. Rev. B}\ }\textbf {\bibinfo {volume} {102}},\ \bibinfo
  {pages} {201105} (\bibinfo {year} {2020})}\BibitemShut {NoStop}%
\bibitem [{\citenamefont {Tran}\ \emph {et~al.}(2019)\citenamefont {Tran},
  \citenamefont {Moody}, \citenamefont {Wu}, \citenamefont {Lu}, \citenamefont
  {Choi}, \citenamefont {Kim}, \citenamefont {Rai}, \citenamefont {Sanchez},
  \citenamefont {Quan}, \citenamefont {Singh}, \citenamefont {Embley},
  \citenamefont {Zepeda}, \citenamefont {Campbell}, \citenamefont {Autry},
  \citenamefont {Taniguchi}, \citenamefont {Watanabe}, \citenamefont {Lu},
  \citenamefont {Banerjee}, \citenamefont {Silverman}, \citenamefont {Kim},
  \citenamefont {Tutuc}, \citenamefont {Yang}, \citenamefont {Macdonald},\ and\
  \citenamefont {Li}}]{2019Evidence}%
  \BibitemOpen
  \bibfield  {author} {\bibinfo {author} {\bibfnamefont {K.}~\bibnamefont
  {Tran}}, \bibinfo {author} {\bibfnamefont {G.}~\bibnamefont {Moody}},
  \bibinfo {author} {\bibfnamefont {F.}~\bibnamefont {Wu}}, \bibinfo {author}
  {\bibfnamefont {X.}~\bibnamefont {Lu}}, \bibinfo {author} {\bibfnamefont
  {J.}~\bibnamefont {Choi}}, \bibinfo {author} {\bibfnamefont {K.}~\bibnamefont
  {Kim}}, \bibinfo {author} {\bibfnamefont {A.}~\bibnamefont {Rai}}, \bibinfo
  {author} {\bibfnamefont {D.~A.}\ \bibnamefont {Sanchez}}, \bibinfo {author}
  {\bibfnamefont {J.}~\bibnamefont {Quan}}, \bibinfo {author} {\bibfnamefont
  {A.}~\bibnamefont {Singh}}, \bibinfo {author} {\bibfnamefont
  {J.}~\bibnamefont {Embley}}, \bibinfo {author} {\bibfnamefont
  {A.}~\bibnamefont {Zepeda}}, \bibinfo {author} {\bibfnamefont
  {M.}~\bibnamefont {Campbell}}, \bibinfo {author} {\bibfnamefont
  {T.}~\bibnamefont {Autry}}, \bibinfo {author} {\bibfnamefont
  {T.}~\bibnamefont {Taniguchi}}, \bibinfo {author} {\bibfnamefont
  {K.}~\bibnamefont {Watanabe}}, \bibinfo {author} {\bibfnamefont
  {N.}~\bibnamefont {Lu}}, \bibinfo {author} {\bibfnamefont {S.~K.}\
  \bibnamefont {Banerjee}}, \bibinfo {author} {\bibfnamefont {K.~L.}\
  \bibnamefont {Silverman}}, \bibinfo {author} {\bibfnamefont {S.}~\bibnamefont
  {Kim}}, \bibinfo {author} {\bibfnamefont {E.}~\bibnamefont {Tutuc}}, \bibinfo
  {author} {\bibfnamefont {L.}~\bibnamefont {Yang}}, \bibinfo {author}
  {\bibfnamefont {A.~H.}\ \bibnamefont {Macdonald}},\ and\ \bibinfo {author}
  {\bibfnamefont {X.}~\bibnamefont {Li}},\ }\href
  {https://doi.org/10.1038/s41586-019-0975-z} {\bibfield  {journal} {\bibinfo
  {journal} {Nature}\ }\textbf {\bibinfo {volume} {567}},\ \bibinfo {pages}
  {71} (\bibinfo {year} {2019})}\BibitemShut {NoStop}%
\bibitem [{\citenamefont {Seyler}\ \emph {et~al.}(2019)\citenamefont {Seyler},
  \citenamefont {Rivera}, \citenamefont {Yu}, \citenamefont {Wilson},
  \citenamefont {Ray}, \citenamefont {Mandrus}, \citenamefont {Yan},
  \citenamefont {Yao},\ and\ \citenamefont {Xu}}]{2019Signatures}%
  \BibitemOpen
  \bibfield  {author} {\bibinfo {author} {\bibfnamefont {K.}~\bibnamefont
  {Seyler}}, \bibinfo {author} {\bibfnamefont {P.}~\bibnamefont {Rivera}},
  \bibinfo {author} {\bibfnamefont {H.}~\bibnamefont {Yu}}, \bibinfo {author}
  {\bibfnamefont {N.}~\bibnamefont {Wilson}}, \bibinfo {author} {\bibfnamefont
  {E.}~\bibnamefont {Ray}}, \bibinfo {author} {\bibfnamefont {D.}~\bibnamefont
  {Mandrus}}, \bibinfo {author} {\bibfnamefont {J.}~\bibnamefont {Yan}},
  \bibinfo {author} {\bibfnamefont {W.}~\bibnamefont {Yao}},\ and\ \bibinfo
  {author} {\bibfnamefont {X.}~\bibnamefont {Xu}},\ }\href
  {https://doi.org/10.1038/s41586-019-0957-1} {\bibfield  {journal} {\bibinfo
  {journal} {Nature}\ }\textbf {\bibinfo {volume} {567}},\ \bibinfo {pages}
  {66-70} (\bibinfo {year} {2019})}\BibitemShut {NoStop}%
\bibitem [{\citenamefont {Yu}\ \emph {et~al.}(2011)\citenamefont {Yu},
  \citenamefont {Qi}, \citenamefont {Bernevig}, \citenamefont {Fang},\ and\
  \citenamefont {Dai}}]{PhysRevB.84.075119}%
  \BibitemOpen
  \bibfield  {author} {\bibinfo {author} {\bibfnamefont {R.}~\bibnamefont
  {Yu}}, \bibinfo {author} {\bibfnamefont {X.~L.}\ \bibnamefont {Qi}}, \bibinfo
  {author} {\bibfnamefont {A.}~\bibnamefont {Bernevig}}, \bibinfo {author}
  {\bibfnamefont {Z.}~\bibnamefont {Fang}},\ and\ \bibinfo {author}
  {\bibfnamefont {X.}~\bibnamefont {Dai}},\ }\href
  {https://doi.org/10.1103/PhysRevB.84.075119} {\bibfield  {journal} {\bibinfo
  {journal} {Phys. Rev. B}\ }\textbf {\bibinfo {volume} {84}},\ \bibinfo
  {pages} {075119} (\bibinfo {year} {2011})}\BibitemShut {NoStop}%
\bibitem [{\citenamefont {Sancho}\ \emph {et~al.}(1985)\citenamefont {Sancho},
  \citenamefont {Sancho}, \citenamefont {Sancho},\ and\ \citenamefont
  {Rubio}}]{Sancho_1985}%
  \BibitemOpen
  \bibfield  {author} {\bibinfo {author} {\bibfnamefont {M.~P.~L.}\
  \bibnamefont {Sancho}}, \bibinfo {author} {\bibfnamefont {J.~M.~L.}\
  \bibnamefont {Sancho}}, \bibinfo {author} {\bibfnamefont {J.~M.~L.}\
  \bibnamefont {Sancho}},\ and\ \bibinfo {author} {\bibfnamefont
  {J.}~\bibnamefont {Rubio}},\ }\href
  {https://doi.org/10.1088/0305-4608/15/4/009} {\bibfield  {journal} {\bibinfo
  {journal} {J. Phys. F: Met. Phys}\ }\textbf {\bibinfo {volume} {15}},\
  \bibinfo {pages} {851} (\bibinfo {year} {1985})}\BibitemShut {NoStop}%
\bibitem [{\citenamefont {Wu}\ \emph {et~al.}(2018)\citenamefont {Wu},
  \citenamefont {Zhang}, \citenamefont {Song}, \citenamefont {Troyer},\ and\
  \citenamefont {Soluyanov}}]{WU2017}%
  \BibitemOpen
  \bibfield  {author} {\bibinfo {author} {\bibfnamefont {Q.}~\bibnamefont
  {Wu}}, \bibinfo {author} {\bibfnamefont {S.}~\bibnamefont {Zhang}}, \bibinfo
  {author} {\bibfnamefont {H.-F.}\ \bibnamefont {Song}}, \bibinfo {author}
  {\bibfnamefont {M.}~\bibnamefont {Troyer}},\ and\ \bibinfo {author}
  {\bibfnamefont {A.~A.}\ \bibnamefont {Soluyanov}},\ }\href
  {https://doi.org/10.1016/j.cpc.2017.09.033} {\bibfield  {journal} {\bibinfo
  {journal} {Comput. Phys. Commun.}\ }\textbf {\bibinfo {volume} {224}},\
  \bibinfo {pages} {405 } (\bibinfo {year} {2018})}\BibitemShut {NoStop}%
\end{thebibliography}

\begin{thebibliography}{12}%
\makeatletter
\providecommand \@ifxundefined [1]{%
 \@ifx{#1\undefined}
}%
\providecommand \@ifnum [1]{%
 \ifnum #1\expandafter \@firstoftwo
 \else \expandafter \@secondoftwo
 \fi
}%
\providecommand \@ifx [1]{%
 \ifx #1\expandafter \@firstoftwo
 \else \expandafter \@secondoftwo
 \fi
}%
\providecommand \natexlab [1]{#1}%
\providecommand \enquote  [1]{``#1''}%
\providecommand \bibnamefont  [1]{#1}%
\providecommand \bibfnamefont [1]{#1}%
\providecommand \citenamefont [1]{#1}%
\providecommand \href@noop [0]{\@secondoftwo}%
\providecommand \href [0]{\begingroup \@sanitize@url \@href}%
\providecommand \@href[1]{\@@startlink{#1}\@@href}%
\providecommand \@@href[1]{\endgroup#1\@@endlink}%
\providecommand \@sanitize@url [0]{\catcode `\\12\catcode `\$12\catcode
  `\&12\catcode `\#12\catcode `\^12\catcode `\_12\catcode `\%12\relax}%
\providecommand \@@startlink[1]{}%
\providecommand \@@endlink[0]{}%
\providecommand \url  [0]{\begingroup\@sanitize@url \@url }%
\providecommand \@url [1]{\endgroup\@href {#1}{\urlprefix }}%
\providecommand \urlprefix  [0]{URL }%
\providecommand \Eprint [0]{\href }%
\providecommand \doibase [0]{https://doi.org/}%
\providecommand \selectlanguage [0]{\@gobble}%
\providecommand \bibinfo  [0]{\@secondoftwo}%
\providecommand \bibfield  [0]{\@secondoftwo}%
\providecommand \translation [1]{[#1]}%
\providecommand \BibitemOpen [0]{}%
\providecommand \bibitemStop [0]{}%
\providecommand \bibitemNoStop [0]{.\EOS\space}%
\providecommand \EOS [0]{\spacefactor3000\relax}%
\providecommand \BibitemShut  [1]{\csname bibitem#1\endcsname}%
\let\auto@bib@innerbib\@empty
\bibitem [{\citenamefont {Tong}\ \emph {et~al.}(2017)\citenamefont {Tong},
  \citenamefont {Yu}, \citenamefont {Zhu}, \citenamefont {Wang}, \citenamefont
  {Xu},\ and\ \citenamefont {Yao}}]{SM2017Topological}%
  \BibitemOpen
  \bibfield  {author} {\bibinfo {author} {\bibfnamefont {Q.}~\bibnamefont
  {Tong}}, \bibinfo {author} {\bibfnamefont {H.}~\bibnamefont {Yu}}, \bibinfo
  {author} {\bibfnamefont {Q.}~\bibnamefont {Zhu}}, \bibinfo {author}
  {\bibfnamefont {Y.}~\bibnamefont {Wang}}, \bibinfo {author} {\bibfnamefont
  {X.}~\bibnamefont {Xu}},\ and\ \bibinfo {author} {\bibfnamefont
  {W.}~\bibnamefont {Yao}},\ }\href {https://doi.org/10.1038/nphys3968}
  {\bibfield  {journal} {\bibinfo  {journal} {Nat. Phys.}\ }\textbf {\bibinfo
  {volume} {13}},\ \bibinfo {pages} {356-362} (\bibinfo {year}
  {2017})}\BibitemShut {NoStop}%
\bibitem [{\citenamefont {Milfeld}\ and\ \citenamefont
  {Wyatt}(1983)}]{SMPhysRevA.27.72}%
  \BibitemOpen
  \bibfield  {author} {\bibinfo {author} {\bibfnamefont {K.~F.}\ \bibnamefont
  {Milfeld}}\ and\ \bibinfo {author} {\bibfnamefont {R.~E.}\ \bibnamefont
  {Wyatt}},\ }\href {https://doi.org/10.1103/PhysRevA.27.72} {\bibfield
  {journal} {\bibinfo  {journal} {Phys. Rev. A}\ }\textbf {\bibinfo {volume}
  {27}},\ \bibinfo {pages} {72} (\bibinfo {year} {1983})}\BibitemShut {NoStop}%
\bibitem [{\citenamefont {G\'omez-Le\'on}\ and\ \citenamefont
  {Platero}(2013)}]{SMPhysRevLett.110.200403}%
  \BibitemOpen
  \bibfield  {author} {\bibinfo {author} {\bibfnamefont {A.}~\bibnamefont
  {G\'omez-Le\'on}}\ and\ \bibinfo {author} {\bibfnamefont {G.}~\bibnamefont
  {Platero}},\ }\href {https://doi.org/10.1103/PhysRevLett.110.200403}
  {\bibfield  {journal} {\bibinfo  {journal} {Phys. Rev. Lett.}\ }\textbf
  {\bibinfo {volume} {110}},\ \bibinfo {pages} {200403} (\bibinfo {year}
  {2013})}\BibitemShut {NoStop}%
\bibitem [{\citenamefont {Goldman}\ and\ \citenamefont
  {Dalibard}(2014)}]{SMPhysRevX.4.031027}%
  \BibitemOpen
  \bibfield  {author} {\bibinfo {author} {\bibfnamefont {N.}~\bibnamefont
  {Goldman}}\ and\ \bibinfo {author} {\bibfnamefont {J.}~\bibnamefont
  {Dalibard}},\ }\href {https://doi.org/10.1103/PhysRevX.4.031027} {\bibfield
  {journal} {\bibinfo  {journal} {Phys. Rev. X}\ }\textbf {\bibinfo {volume}
  {4}},\ \bibinfo {pages} {031027} (\bibinfo {year} {2014})}\BibitemShut
  {NoStop}%
\bibitem [{\citenamefont {Hohenberg}\ and\ \citenamefont
  {Kohn}(1964)}]{SMPhysRev.136.B864}%
  \BibitemOpen
  \bibfield  {author} {\bibinfo {author} {\bibfnamefont {P.}~\bibnamefont
  {Hohenberg}}\ and\ \bibinfo {author} {\bibfnamefont {W.}~\bibnamefont
  {Kohn}},\ }\href {https://doi.org/10.1103/PhysRev.136.B864} {\bibfield
  {journal} {\bibinfo  {journal} {Phys. Rev.}\ }\textbf {\bibinfo {volume}
  {136}},\ \bibinfo {pages} {B864} (\bibinfo {year} {1964})}\BibitemShut
  {NoStop}%
\bibitem [{\citenamefont {Kohn}\ and\ \citenamefont
  {Sham}(1965)}]{SMPhysRev.140.A1133}%
  \BibitemOpen
  \bibfield  {author} {\bibinfo {author} {\bibfnamefont {W.}~\bibnamefont
  {Kohn}}\ and\ \bibinfo {author} {\bibfnamefont {L.~J.}\ \bibnamefont
  {Sham}},\ }\href {https://doi.org/10.1103/PhysRev.140.A1133} {\bibfield
  {journal} {\bibinfo  {journal} {Phys. Rev.}\ }\textbf {\bibinfo {volume}
  {140}},\ \bibinfo {pages} {A1133} (\bibinfo {year} {1965})}\BibitemShut
  {NoStop}%
\bibitem [{\citenamefont {Kresse}\ and\ \citenamefont
  {Furthm\"uller}(1996)}]{SMPhysRevB.54.11169}%
  \BibitemOpen
  \bibfield  {author} {\bibinfo {author} {\bibfnamefont {G.}~\bibnamefont
  {Kresse}}\ and\ \bibinfo {author} {\bibfnamefont {J.}~\bibnamefont
  {Furthm\"uller}},\ }\href {https://doi.org/10.1103/PhysRevB.54.11169}
  {\bibfield  {journal} {\bibinfo  {journal} {Phys. Rev. B}\ }\textbf {\bibinfo
  {volume} {54}},\ \bibinfo {pages} {11169} (\bibinfo {year}
  {1996})}\BibitemShut {NoStop}%
\bibitem [{\citenamefont {Perdew}\ \emph {et~al.}(1996)\citenamefont {Perdew},
  \citenamefont {Burke},\ and\ \citenamefont
  {Ernzerhof}}]{SMPhysRevLett.77.3865}%
  \BibitemOpen
  \bibfield  {author} {\bibinfo {author} {\bibfnamefont {J.~P.}\ \bibnamefont
  {Perdew}}, \bibinfo {author} {\bibfnamefont {K.}~\bibnamefont {Burke}},\ and\
  \bibinfo {author} {\bibfnamefont {M.}~\bibnamefont {Ernzerhof}},\ }\href
  {https://doi.org/10.1103/PhysRevLett.77.3865} {\bibfield  {journal} {\bibinfo
   {journal} {Phys. Rev. Lett.}\ }\textbf {\bibinfo {volume} {77}},\ \bibinfo
  {pages} {3865} (\bibinfo {year} {1996})}\BibitemShut {NoStop}%
\bibitem [{\citenamefont {Burns}\ \emph {et~al.}()\citenamefont {Burns},
  \citenamefont {Mayagoitia}, \citenamefont {Álvaro Vázquez}, \citenamefont
  {Sumpter}, \citenamefont {G.}, \citenamefont {Sherrill},\ and\ \citenamefont
  {David}}]{SMDensity}%
  \BibitemOpen
  \bibfield  {author} {\bibinfo {author} {\bibfnamefont {L.~A.}\ \bibnamefont
  {Burns}}, \bibinfo {author} {\bibnamefont {Mayagoitia}}, \bibinfo {author}
  {\bibnamefont {Álvaro Vázquez}}, \bibinfo {author} {\bibnamefont
  {Sumpter}}, \bibinfo {author} {\bibfnamefont {B.}~\bibnamefont {G.}},
  \bibinfo {author} {\bibnamefont {Sherrill}},\ and\ \bibinfo {author}
  {\bibfnamefont {C.}~\bibnamefont {David}},\ }\href
  {https://doi.org/10.1063/1.3545971} {\bibfield  {journal} {\bibinfo
  {journal} {J. Chem. Phys.}\ }\textbf {\bibinfo {volume} {134}},\ \bibinfo
  {pages} {084107}}\BibitemShut {NoStop}%
\bibitem [{\citenamefont {Mostofi}\ \emph {et~al.}(2014)\citenamefont
  {Mostofi}, \citenamefont {Yates}, \citenamefont {Pizzi}, \citenamefont {Lee},
  \citenamefont {Souza}, \citenamefont {Vanderbilt},\ and\ \citenamefont
  {Marzari}}]{SMMostofi2014}%
  \BibitemOpen
  \bibfield  {author} {\bibinfo {author} {\bibfnamefont {A.~A.}\ \bibnamefont
  {Mostofi}}, \bibinfo {author} {\bibfnamefont {J.~R.}\ \bibnamefont {Yates}},
  \bibinfo {author} {\bibfnamefont {G.}~\bibnamefont {Pizzi}}, \bibinfo
  {author} {\bibfnamefont {Y.-S.}\ \bibnamefont {Lee}}, \bibinfo {author}
  {\bibfnamefont {I.}~\bibnamefont {Souza}}, \bibinfo {author} {\bibfnamefont
  {D.}~\bibnamefont {Vanderbilt}},\ and\ \bibinfo {author} {\bibfnamefont
  {N.}~\bibnamefont {Marzari}},\ }\href
  {https://doi.org/https://doi.org/10.1016/j.cpc.2014.05.003} {\bibfield
  {journal} {\bibinfo  {journal} {Comput. Phys. Commun.}\ }\textbf {\bibinfo
  {volume} {185}},\ \bibinfo {pages} {2309 } (\bibinfo {year}
  {2014})}\BibitemShut {NoStop}%
\bibitem [{\citenamefont {Sancho}\ \emph {et~al.}(1985)\citenamefont {Sancho},
  \citenamefont {Sancho}, \citenamefont {Sancho},\ and\ \citenamefont
  {Rubio}}]{SMSancho_1985}%
  \BibitemOpen
  \bibfield  {author} {\bibinfo {author} {\bibfnamefont {M.~P.~L.}\
  \bibnamefont {Sancho}}, \bibinfo {author} {\bibfnamefont {J.~M.~L.}\
  \bibnamefont {Sancho}}, \bibinfo {author} {\bibfnamefont {J.~M.~L.}\
  \bibnamefont {Sancho}},\ and\ \bibinfo {author} {\bibfnamefont
  {J.}~\bibnamefont {Rubio}},\ }\href
  {https://doi.org/10.1088/0305-4608/15/4/009} {\bibfield  {journal} {\bibinfo
  {journal} {J. Phys. F: Met. Phys}\ }\textbf {\bibinfo {volume} {15}},\
  \bibinfo {pages} {851} (\bibinfo {year} {1985})}\BibitemShut {NoStop}%
\bibitem [{\citenamefont {Wu}\ \emph {et~al.}(2018)\citenamefont {Wu},
  \citenamefont {Zhang}, \citenamefont {Song}, \citenamefont {Troyer},\ and\
  \citenamefont {Soluyanov}}]{SMWU2017}%
  \BibitemOpen
  \bibfield  {author} {\bibinfo {author} {\bibfnamefont {Q.}~\bibnamefont
  {Wu}}, \bibinfo {author} {\bibfnamefont {S.}~\bibnamefont {Zhang}}, \bibinfo
  {author} {\bibfnamefont {H.-F.}\ \bibnamefont {Song}}, \bibinfo {author}
  {\bibfnamefont {M.}~\bibnamefont {Troyer}},\ and\ \bibinfo {author}
  {\bibfnamefont {A.~A.}\ \bibnamefont {Soluyanov}},\ }\href
  {https://doi.org/10.1016/j.cpc.2017.09.033} {\bibfield  {journal} {\bibinfo
  {journal} {Comput. Phys. Commun.}\ }\textbf {\bibinfo {volume} {224}},\
  \bibinfo {pages} {405 } (\bibinfo {year} {2018})}\BibitemShut {NoStop}%
\end{thebibliography}

%

\newpage

\begin{widetext}
\newpage

\setcounter{figure}{0}
\setcounter{equation}{0}
\makeatletter

\makeatother
\renewcommand{\thefigure}{S\arabic{figure}}
\renewcommand{\thetable}{S\Roman{table}}
\renewcommand{\theequation}{S\arabic{equation}}

\begin{center}
	\textbf{
		\large{Supplemental Material for}}
	\vspace{0.2cm}
	
	\textbf{
		\large{
			``Floquet Valley-Polarized Quantum Anomalous Hall State in Nonmagnetic Heterobilayers"}
	}
\end{center}

In this Supplemental Material, we give detailed derivation of the low-energy effective Hamiltonian based on the Floquet theorem, computational methods of first-principles calculations, lattice structures and electronic band structures for a series of potential TMD heterobilayers, implementation of the Floquet theorem in Wannier-function-based tight-binding (WFTB) model, and optically switchable topological properties dependent on handedness.

\section{Effective low energy time-independent Hamiltonian}

Two massive Dirac materials can hybidize through interlayer hopping, which can be described by the minimal $\rm{\mathbf{k \cdot p}}$ model \cite{SM2017Topological}:
\begin{equation}\label{EqS1}
H_{\tau}(\mathbf{k})=
\begin{pmatrix}
-\mathit{\Delta}/2+M_{u}          &{\nu}_{u}(\tau  k_{x}-i\epsilon k_{y})      & t_{cc}              & t_{cv} \\
{\nu}_{u}(\tau k_{x}+i\epsilon k_{y})      & -\mathit{\Delta}/2                         &  t_{vc}             & t_{vv} \\
t^{*}_{cc}           & t^{*}_{vc}           &\mathit{\Delta}/2                        &{\nu}_{l}(\tau k_{x}-ik_{y}) \\
t^{*}_{cv}          & t^{*}_{vv}           &{\nu}_{l}(\tau k_{x}+ik_{y})          &\mathit{\Delta}/2-M_{l}
\end{pmatrix},
\end{equation}
in which $\nu_{u}(\nu_{l})$ and $M_{u}(M_{l})$ corresponds to Fermi velocity and mass in the upper (lower) layer. $\tau=\pm 1$ is the valley index of hexagonal bilayer. $\epsilon=\pm 1$ denotes the two types of stacking orientations, $\epsilon=+1$ ($\epsilon=-1$) corresponds to R-type (H-type) stacking in the main text. $t_{ij}(i,j=c,v)$ are interlayer hopping between band i of upper layer and the one in band j of lower layer at the $K/K{'}$ valley. The $\mathit{\Delta}=\mathit{\Delta}_{g}-U$ is the heterobilayer band gap under external electric field, where $U$ is the interlayer bias proportional to the perpendicular electric field, and $\mathit{\Delta}_{g}$ is heterobilayer intrinsic band gap. It is worth noting that certain interlayer hopping channels must vanish considering the three-fold rotational symmetry $C_3$ of hexagonal bilayer \cite{SM2017Topological}. Herein, we choose $H_X^M$ stacking with the characteristics of valley quantum spin Hall (VQSH) state as a representative example to elaborate the topological phases. Consequently, the Eq. \ref{EqS1} becomes
\begin{equation}\label{EqS2}
\begin{split}
H_{\tau}(\mathbf{k})&= \boldsymbol s_x
(\tau v_1k_{x}\boldsymbol\sigma_{x}+ v_2 k_{y}\boldsymbol\sigma_{y}) +
\boldsymbol s_y
( v_1k_{y}\boldsymbol\sigma_{x}- \tau v_2 k_{x}\boldsymbol\sigma_{y}) \\
& +\frac{\Delta}{2}\boldsymbol s_z\boldsymbol\sigma_{z} +M_2(\boldsymbol s_z\boldsymbol\sigma_{0}-\boldsymbol s_0\boldsymbol\sigma_{0}) +
  M_1(\boldsymbol s_0\boldsymbol\sigma_{z}-\boldsymbol s_z\boldsymbol\sigma_{z}) \\
&  + \lambda(\boldsymbol s_x\boldsymbol\sigma_{0} - \boldsymbol s_x\boldsymbol\sigma_{z}),
\end{split}
\end{equation}
in which $v_{1,2} = {(v_l \pm v_u)}/{2}$, $M_{1,2} = {(M_l \pm M_u)}/{4}$, and $\lambda=\frac{1}{2}t_{vv}$ denotes interlayer hopping for holes. In this case, the interlayer hopping for electrons vanishes at $K$ points. The gap-closing topological phase transition regulated by the electric field occurs when $\mathit{\Delta}$ is varied from $\mathit{\Delta}>0$ to $\mathit{\Delta}<0$.  Here we only focus on neighborhood regime around the critical point $\mathit{\Delta}\approx 0$, i.e., $M_{u}(M_{l}) \gg \mathit{\Delta}$. In this case, the $4\times4$ Hamiltonian can be projected to a $2\times2$ Hamiltonian:
\begin{equation}\label{EqS3}
\begin{split}
H_{\tau}(\mathbf{k}) &\cong (\frac{\mathit{\Delta}}{2}+\frac{\mathit{P}}{2}k^2)\boldsymbol{\sigma}_{z}+\frac{\mathit{Q}}{2}k^2+2\lambda\frac{\nu_{l}}{M_{l}} (\tau k_{x} - ik_{y})(\boldsymbol{\sigma}_{x}+i\boldsymbol{\sigma}_{y})+h.c. \\
&= d_{0}(k)\sigma_{0}+\mathbf{d_{\tau}(k)} \cdot \boldsymbol{\sigma},
\end{split}
\end{equation}
where $\mathit{P}=\nu^{2}_{u}/M_{u}+\nu^{2}_{l}/M_{l}$ and $\mathit{Q}=\nu^{2}_{u}/M_{u}-\nu^{2}_{l}/M_{l}$. We have $d_{0}(k) = -Dk^2$ and $\mathbf{d_{\tau}(k)}=(\tau Ck_{x}, Ck_{y}, M-Bk^{2})$, where $D=-\frac{Q}{2}$, $C=4\lambda \frac{\nu_{l}}{M_{l}}$, $M=\frac{\mathit{\Delta}}{2}$, and $B=-\frac{P}{2}$.

Based on the Floquet theorem, under the irradiation of time-periodic circularly polarized light (CPL) with vector potential of $\mathbf{A}(t)=\mathbf{A}({t}+T)= A[\cos (\omega \tau ), \pm \eta \sin (\omega \tau ), 0]$, where $\eta=\pm 1$ denotes the left(right)-handed CPL. The $A$, $\omega$, and $T=2\pi /\omega$ are its amplitude, frequency and time period, respectively. The vector potential $\mathbf{A}(t)$ can be coupled to Hamiltonian based on the minimal coupling substitution, $H_{\tau}(\mathbf{k},t)=H_{\tau}(\mathbf{k}+e\mathbf{A(t)})$ \cite{SMPhysRevA.27.72,SMPhysRevLett.110.200403}. The time-periodic Hamiltonian can be expanded as
\begin{equation}\label{EqS4}
H_{\tau}(\mathbf{k},t)=\sum_{m}H_{\tau}^{m}(\mathbf{k})e^{im\omega t},
\end{equation}
and then the effective Floquet Hamiltonian in the high frequency approximation can be expressed as \cite{SMPhysRevX.4.031027}
\begin{equation}\label{EqS5}
H_{\tau}^{F}(\mathbf{k}) = H^{0}(\mathbf{k})+\sum_{m\ge 1}\frac{\left[H_{\tau}^{-m},H_{\tau}^{m}\right]}{m\hbar \omega},
\end{equation}
with
\begin{equation}
\begin{split}\label{EqS6}
&H^{0}(\mathbf{k}) = \frac{Q}{2}[(\frac{eA}{\hbar})^{2}+k^{2}]\sigma_{0}+4\tau\lambda\frac{\nu_{l}}{M_{l}} k_{x}\sigma_{x}+4\lambda\frac{\nu_{l}}{M_{l}}k_{y}\sigma_{y}+[\frac{P}{2}(\frac{eA}{\hbar})^{2}+\frac{P}{2}k^{2}+\frac{\mathit{\Delta}}{2}]\sigma_{z}, \\
&H^{\pm1}(\mathbf{k}) = \frac{Q}{2}\frac{eA}{\hbar}(k_{x} \pm i\eta k_{y})\sigma_{0}+2\tau \lambda P\frac{eA}{\hbar}\frac{\nu_{l}}{M_{l}}\sigma_{x}\pm 2i\eta\lambda\frac{eA}{\hbar} \frac{\nu_{l}}{M_{l}}\sigma_{y}+\frac{P}{2}\frac{eA}{\hbar}(k_{x} \pm i\eta k_{y})\sigma_{z}, \\
&H^{\pm2}(\mathbf{k}) = 0.
\end{split}
\end{equation}
By directly calculating the commutators,
\begin{equation}
\begin{split}\label{EqS7}
[H^{-1},H^{1}] &=  -4\eta\lambda\frac{\nu_{l}}{M_{l}}P(\frac{eA}{\hbar})^{2}k_{x}\sigma_{x} - 4\eta\tau\lambda\frac{\nu_{l}}{M_{l}}P(\frac{eA}{\hbar})^{2}k_{y}\sigma_{y}+8\lambda(\frac{\nu_{l}}{M_{l}})^{2}(\frac{eA}{\hbar})^{2}\sigma_{z} \\
[H^{-2},H^{2}] &= 0
\end{split}
\end{equation}
we obtain
\begin{equation}\label{EqS8}
H_{\tau}^{F}(\mathbf{k}) = \tilde{d_{0}}(k)\sigma_{0}+\tilde{\mathbf{d}}_{\tau}(\mathbf{k}) \cdot \boldsymbol{\sigma},
\end{equation}
the light renormalized $\tilde{d_{0}}(k) = \tilde{D}-Dk^2$ and $\tilde{\mathbf{d}}_{\tau}(\mathbf{k})=(\tau \tilde{C}k_{x}, \tilde{C}k_{y}, \tilde{M}_{\tau}-Bk^{2})$, where $\tilde{D}=-D(\frac{eA}{\hbar})^2$, $\tilde{C}=C-\frac{4\eta}{\hbar \omega}\frac{\nu_{l}}{M_{l}}P\left(\frac{eA}{\hbar}\right)^{2}\lambda$, and $\tilde{M}_{\tau}=\frac{\mathit{\Delta}}{2}+(\frac{eA}{\hbar})^{2}[\frac{P}{2}+\frac{8\eta\tau}{\hbar \omega}(\frac{\nu_{l}}{M_{l}})^{2}\lambda]$. As shown in Fig. \ref{Fig. S1}, the Dirac mass $\tilde{M}_{\tau}$ is valley-dependent, leading to two valley sectors with different responses to CPL. The valley-resolved Chern number can be analytically obtained as $C_{\tau}=-\frac{\tau}{2}[\mathrm{sgn}(\tilde{M}_{\tau}) + \mathrm{sgn}(B)]$.

\begin{figure}[H]
    \centering
    \includegraphics[width=12cm]{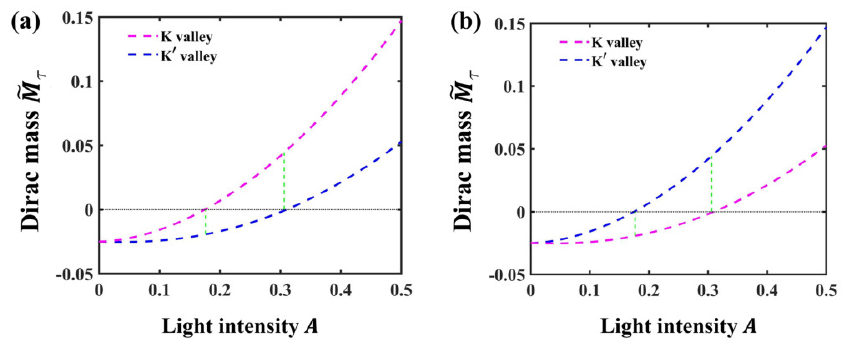}
    \caption{The light-renormalized Dirac mass $\tilde{M}_{\tau}$ as function of light intensity $A$ for (a) left-handed and (b) right-handed CPL. Here, $\mathit{\Delta}=-0.05$, $\lambda=1.5$, and $\frac{\nu_{l}}{M_{l}}=\frac{\nu_{u}}{M_{u}}=0.5$ due to the similar band structures in lower and upper layers for heterobilayers.
    }
    \label{Fig. S1}
\end{figure}

\section{Computational methods}

The first-principles calculations were performed within the framework of density functional theory (DFT) \cite{SMPhysRev.136.B864,SMPhysRev.140.A1133} using the projector augmented-wave method encoded in the Vienna $ab$ $initio$ Simulation Package (VASP) \cite{SMPhysRevB.54.11169}. The exchange correlation functional was described by using the generalized gradient approximation within the Perdew-Burke-Ernzerhof (GGA-PBE) formalism  \cite{SMPhysRevLett.77.3865}. The plane-wave cutoff energy was set to be 500 eV. The first Brillouin zone was sampled by $15 \times 15 \times 1$ Monkhorst-Pack mesh grid. We adopted $1 \times 1$ primitive cell of two different 2H-TMDs to build the heterobilayer. A vacuum layer larger than 20\text{\AA} was introduced to ensure decouple the interaction between two neighboring heterobilayer slabs. All geometric structures are fully relaxed until energy and force converged to $10^{-6}$ eV and 0.01 eV/\text{\AA}. Weak van der Waals (vdW) interactions were included in our calculations using Grimme (DFT-D3) method \cite{SMDensity}. The 56 projected atomic orbitals comprised of $s$ and $d$ orbitals centered at the metal atoms, $s$ and $p$ orbitals centered at the chalcogen atoms in the primitive unit cell were used to construct WFTB model based on maximally localized Wannier functions methods by the WANNIER90 package \cite{SMMostofi2014}. The iterative Green's function method \cite{SMSancho_1985} as implemented in the WannierTools \cite{SMWU2017} was used for chiral edge states calculations.

\section{Lattice structures and stacking-dependent electronic band structures}

\begin{table}[H]
\caption{Structure properties of $H_X^M$ stacking in different heterobilayer. The $a$, $\delta$, $d$, $E_{b}$, and $\Delta\Phi$ represent the optimized lattice constants, lattice mismatch, averaged layer distances, binding energies and work function differences between two isolated monolayers, respectively.}
\begin{center}
\renewcommand\arraystretch{1.5}
\begin{tabular}{p{2.5cm}<{\centering} p{2cm}<{\centering} p{2cm}<{\centering} p{2cm}<{\centering} p{2cm}<{\centering} p{2cm}<{\centering}}
		\hline
\hline
		 & $a(\text{\AA})$ & $\delta(\%)$ & $d(\text{\AA})$ & $E_{b}$(eV) & $\Delta\Phi$(eV)\\
       \hline
		MoS$_2$/WTe$_2$ & 3.33 & 4.50 & 3.21 & -0.30 & 1.32\\
		
		WS$_2$/WTe$_2$ & 3.33 & 4.20 & 3.21 & -0.31 & 1.06\\
		
		MoSe$_2$/WTe$_2$ & 3.41 & 2.63 & 3.24 & -0.31 & 0.63\\
		
		WSe$_2$/WTe$_2$ & 3.40 & 2.35 & 3.24 & -0.32 & 0.49\\

           MoTe$_2$/WTe$_2$ & 3.52 & 0.85 & 3.39 & -0.33 & 0.26\\
		\hline
\hline
\end{tabular}
\end{center}\label{Tab. S1}
\end{table}

\begin{figure}[H]
    \centering
    \includegraphics[width=8.6cm]{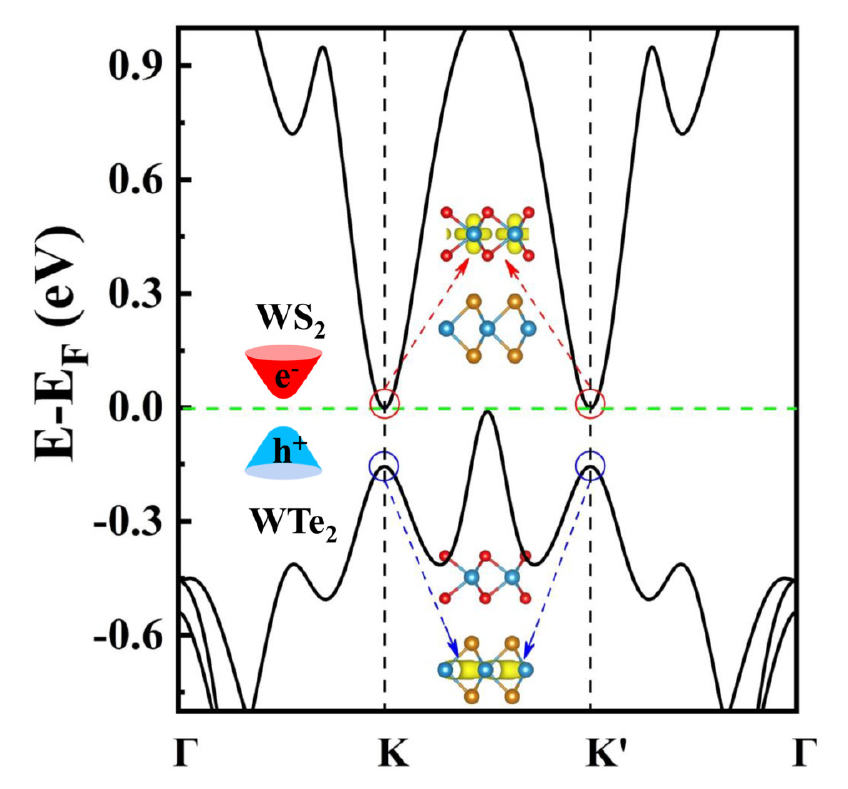}
    \caption{The WS$_2$/WTe$_2$ heterobilayer band structure without SOC. The illustration depicts the type-\uppercase\expandafter{\romannumeral2} band alignment, where  holes(electrons) in two valleys reside in lower (upper) layer.}
    \label{Fig. S2}
\end{figure}

\begin{figure}[H]
    \centering
    \includegraphics[width=17.3cm]{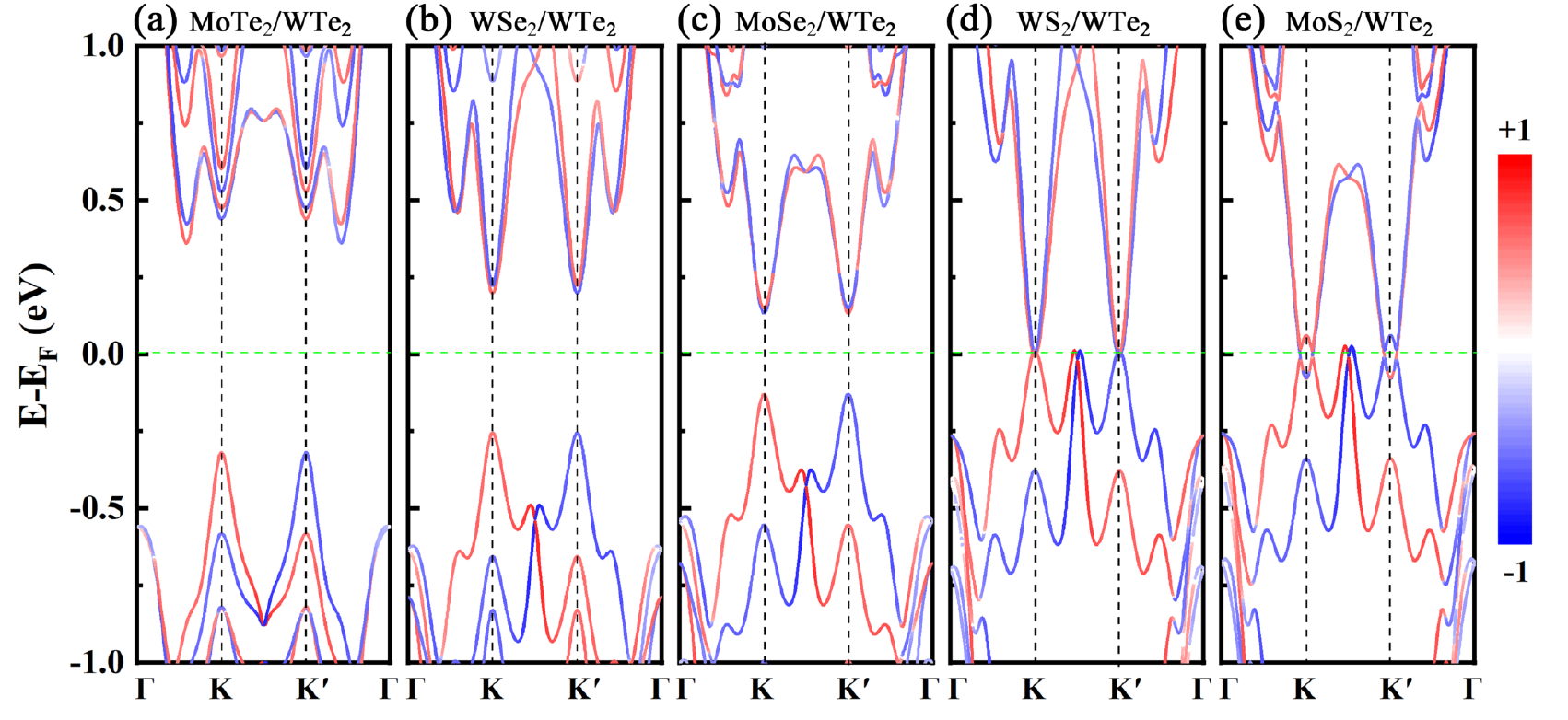}
    \caption{(a)-(e) The spin-resolved band structures for a series of potential TMD heterobilayers in the presence of SOC. Red (blue) color denotes the spin-up (spin-down) component.}
    \label{Fig. S3}
\end{figure}

The Fig. \ref{Fig. S2} shows that holes(electrons) at two valleys are predominantly localized in an individual layer, exhibiting type-\uppercase\expandafter{\romannumeral2} band alignment features. These separation of electrons and holes can effectively bind together to form interlayer excitons. Here, without loss of generality, we started from a series of potential TMD heterobilayers of MX$_2$/WTe$_2$ (M=Mo, W; X= S, Se) in Tab. \ref{Tab. S1}. The choice of these candidates was based on their work function differences between the upper and lower layers. As illustrated in Fig. \ref{Fig. S3}, the calculated results confirmed that band gaps of heterobilayers decrease with increasing the work function differences. To obtain band inversion in the presence of SOC, the large work function difference of two layers is preferred.

\begin{figure}[H]
    \centering
    \includegraphics[width=15cm]{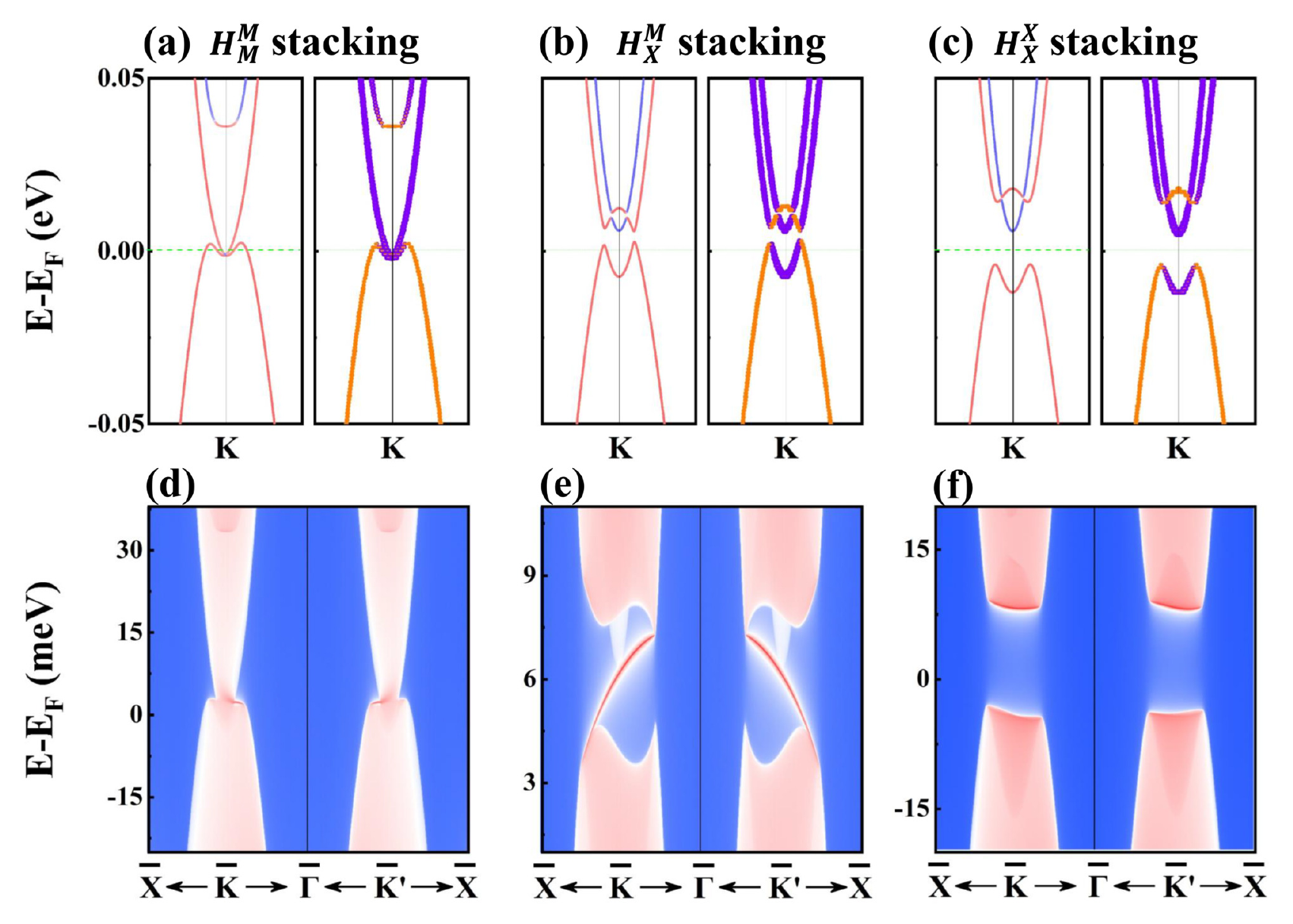}
    \caption{(a)-(c) The enlarged views of spin (left panel) and orbit (right panel) components at the $K$ valley for different high-symmetry stacking order. The component of W-$d_{z^{2}}$ orbital of upper WS$_2$ layer (W-$d_{xy}$\&$d_{x^{2}-{y^{2}}}$ orbital of lower WTe$_2$ layer) is proportional to the width of the purple (orange) curve.  (d)-(f) Corresponding semi-infinite local density of states (LDOS).}
    \label{Fig. S4}
\end{figure}
\begin{figure}[H]
    \centering
    \includegraphics[width=15cm]{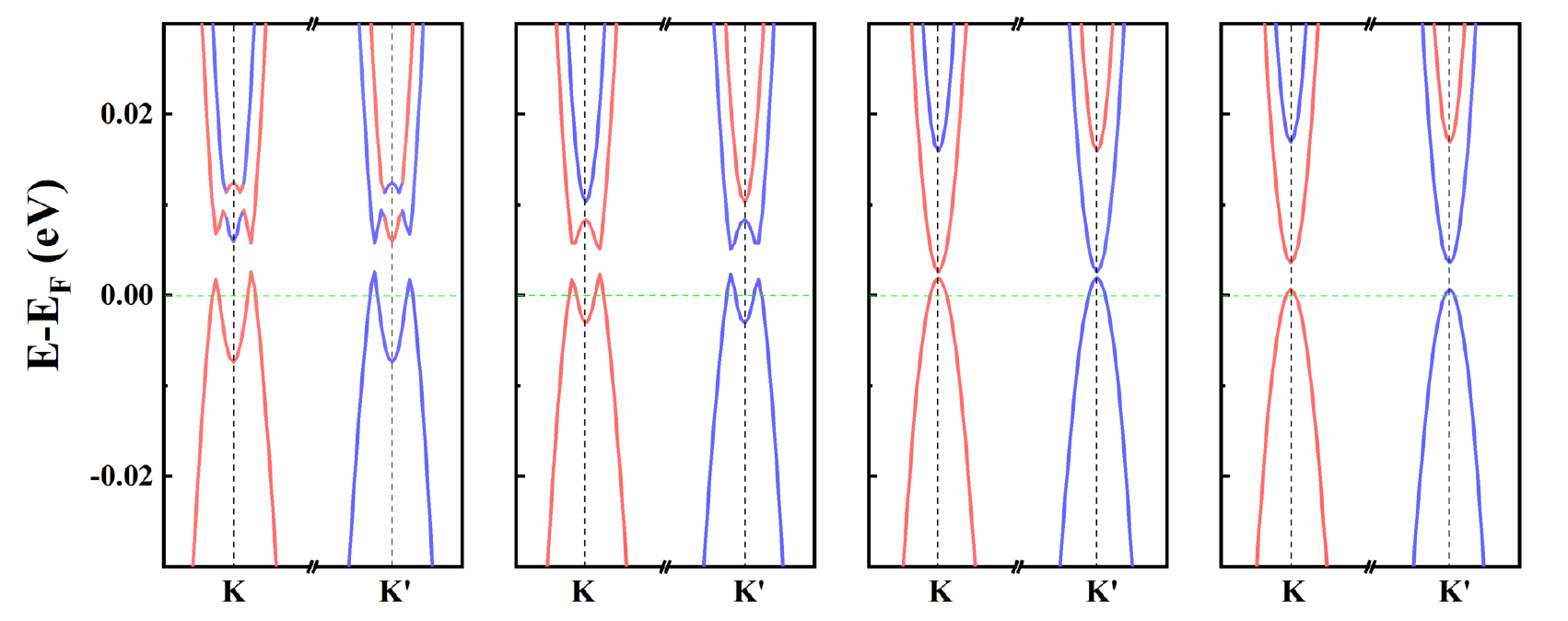}
    \caption{For $H_X^M$ stacking, the band structure evolution as external E-field strength at $K$/$K'$ valleys. From left to right panel, the external E-field strength is set to 0, -0.03,-0.09 and -0.1 V/\AA, respectively. Red (blue) color denotes the spin-up (spin-down) component.}
    \label{Fig. S5}
\end{figure}

\section{Implementation of the Floquet theorem in Wannier-function-based tight-binding (WFTB) model}

Starting from the real-space Wannier tight-binding (TB) Hamiltonian
\begin{equation}\label{Eq.9}
H(\mathbf{r})=\sum_{mn}\sum_{j}t_{j}^{mn}C_{m}^{\dag}(\mathbf{R}_{j})C_{n}(\mathbf{R}_{0})+h.c.,
\end{equation}
where $t_{j}^{mn}$ is a hopping parameter from orbital $n$ located at $\mathbf{R}_{0}$ to orbital $m$ located at $\mathbf{R}_{j}$, and $C_m^{\dag}(\mathbf{R}_{j})$ ($C_m (\mathbf{R}_{0})$) creates (annihilates) an electron on site $\mathbf{R}_{j}$ ($\mathbf{R}_{0}$). We consider it under the
irradiation of an external time-periodic CPL with vector potential of $\mathbf{A}(t)=\mathbf{A}(t+T)= A[\cos (\omega t ), \pm \eta \sin (\omega t ), 0]$. The time-dependent hopping parameter $t_j^{mn}(t)$ is coupled to the Hamiltonian by introducing Peierls substitution \cite {SMPhysRevA.27.72,SMPhysRevLett.110.200403}
\begin{equation}\label{Eq.10}
t_j^{mn}(t) =t_j^{mn}e^{i\frac{e}{\hbar}\mathbf{A}(t)\cdot \mathbf{d}_{mn}},
\end{equation}
where $\mathbf{d}_{mn}$ is the related position vector between two Wannier orbitals. This time-dependent tight-binding Hamiltonian can be written as
\begin{equation}\label{Eq.11}
H(\mathbf{k},t)=\sum_{mn}\sum_{j}t_j^{mn}(t)e^{i\mathbf{k}\cdot\mathbf{R}_j}C_{m}^{\dag}(\mathbf{k}, t)C_{n}(\mathbf{k}, t)+h.c.,
\end{equation}
Taking into account the lattice and time translation invariance, the time-dependent Hamiltonian can be effectively treated with Floquet theorem by performing dual the Fourier transformation. Here, the Floquet-Bloch creation (annihilation) operator can be expressed as
\begin{equation}\label{Eq.12}
\begin{split}
C_{m}^{\dag}(\mathbf{k}, t)&=\sum_{j}\sum_{\alpha=-\infty}^{\infty} C_{\alpha m}^{\dag}(\mathbf{R}_{j})e^{+i\mathbf{k}\cdot \mathbf{R}_{j} -  i\alpha \omega t},  \\
C_{m}(\mathbf{k}, t)&=\sum_{j}\sum_{\alpha=-\infty}^{\infty} C_{\alpha m}(\mathbf{R}_{j})e^{-i\mathbf{k}\cdot \mathbf{R}_{j} +  i\alpha \omega t}.
\end{split}
\end{equation}
Plugging Eq. \ref{Eq.12} into Eq. \ref{Eq.11}, we can obtain an effective static Hamiltonian in the frequency and momentum space
\begin{equation}\label{Eq.13}
H({\mathbf{k}}, \omega)=\sum_{m, n}\sum_{\alpha, \beta}[H_{mn}^{\alpha-\beta}({\mathbf{k}}, \omega)
+\alpha \hbar \omega \delta_{mn}\delta_{\alpha \beta}]C_{\alpha m}^{\dag}(\mathbf{k})C_{\beta n}(\mathbf{k})+h.c,
\end{equation}
where $\hbar \omega$ represents the energy of photo, $(\alpha, \beta)$ is the Floquet index ranging from $-\infty$ to $+\infty$, and the matrix element $H_{mn}^{\alpha-\beta}({\mathbf{k}}, \omega)$ is
\begin{equation}\label{Eq.14}
H_{mn}^{\alpha-\beta}({\mathbf{k}}, \omega)=\sum_{j} e^{i\mathbf{k} \cdot \mathbf{R}_j}\bigg(\frac{1}{T}\int_{0}^{T}t_j^{mn}\times e^{i\frac{e}{\hbar}\mathbf{A}(t)\cdot \mathbf{d}_{mn}}e^{i(\alpha-\beta)\omega t}dt \bigg).
\end{equation}
We give $\mathbf{A}(t)$ and $\mathbf{d}_{mn}$ as the following general forms
\begin{equation}\label{Eq.15}
\mathbf{A}(t) = [A_{x}\sin{(\omega t + {\varphi_1})},A_{y}\sin{(\omega t + {\varphi_2})},A_{z}\sin{(\omega t + {\varphi_3})}],
\end{equation}
\begin{equation}\label{Eq.16}
{\mathbf{d}_{mn}} = (d_{x},d_{y},d_{z}),
\end{equation}
Then, the the matrix element ${H_{mn}^q} (\mathbf{k}, \omega)$ ($q=\alpha-\beta$) can rewrite as
\begin{equation}\label{Eq.17}
H_{mn}^q ({\mathbf{k}}, \omega)=\sum_{j} t_j^{mn}e^{i\mathbf{k}\cdot \mathbf{R}_j}\cdot J_{q}(\frac{e}{\hbar}A_{max}),
\end{equation}
in which $J_q$ is q-th Bessel function
\begin{equation}\label{Eq.18}
J_{q}(\frac{e}{\hbar}A_{max}) = e^{-iq \varphi} \frac{1}{T} \int_{0}^{T} e^{i[\frac{e}{\hbar}A_{max}\sin{(\omega t + \varphi)]}}e^{iq (\omega t+\varphi)}dt ,
\end{equation}
with
\begin{gather}\label{Eq.19}
A_{max} = \sqrt{(A_{x}d_{x}\sin{\varphi _1} + A_{y}d_{y}\sin{\varphi _2}+A_{z}d_{z}\sin{\varphi _3})^2 + (A_{x}d_{x}\cos{\varphi _1} + A_{y}d_{y}\cos{\varphi _2}+A_{z}d_{z}\cos{\varphi _3})^2}, \nonumber \\
\varphi  =  \arctan {(\frac{A_{x}d_{x}\sin{\varphi _1} + A_{y}d_{y}\sin{\varphi _2}+A_{z}d_{z}\sin{\varphi _3}} {A_{x}d_{x}\cos{\varphi _1} + A_{y}d_{y}\cos{\varphi _2}+A_{z}d_{z}\cos{\varphi _3}})}.
\end{gather}
The Floquet-Bloch Hamiltonian in the block matrix form
\begin{equation}\label{Eq.20}
H({\mathbf{k}}, \omega)=
\begin{pmatrix}
\ddots   & \vdots   & \vdots   & \vdots   & \begin{sideways}$\ddots$\end{sideways}   \\
\cdots   & H_{0}-\hbar \omega  & H_{-1}   & H_{-2}  & \cdots  \\
\cdots   & H_{1}  & H_{0}   & H_{-1}  & \cdots \\
\cdots   & H_{2}  & H_{1}  &H_{0}+\hbar \omega  & \cdots \\
\begin{sideways}$\ddots$\end{sideways}   & \vdots   & \vdots   & \vdots   & \ddots   \\
\end{pmatrix}
\end{equation}

As shown in equation (\ref {Eq.20}), the infinite Floquet energy subbands can be truncated at first order $(\left| \alpha - \beta \right|=0,1)$, which is sufficient to obtaining desired convergence as shown in Fig. \ref {Fig. S6}. Photon energy $\hbar \omega = 15$ eV is chosen to be larger than the bandwidth so that overlap between different Floquet subbands is negligible. As a result, the Floquet bands near the Fermi energy are determined by leading order $H_0$.

\begin{figure}[H]
    \centering
    \includegraphics[width=8.6cm]{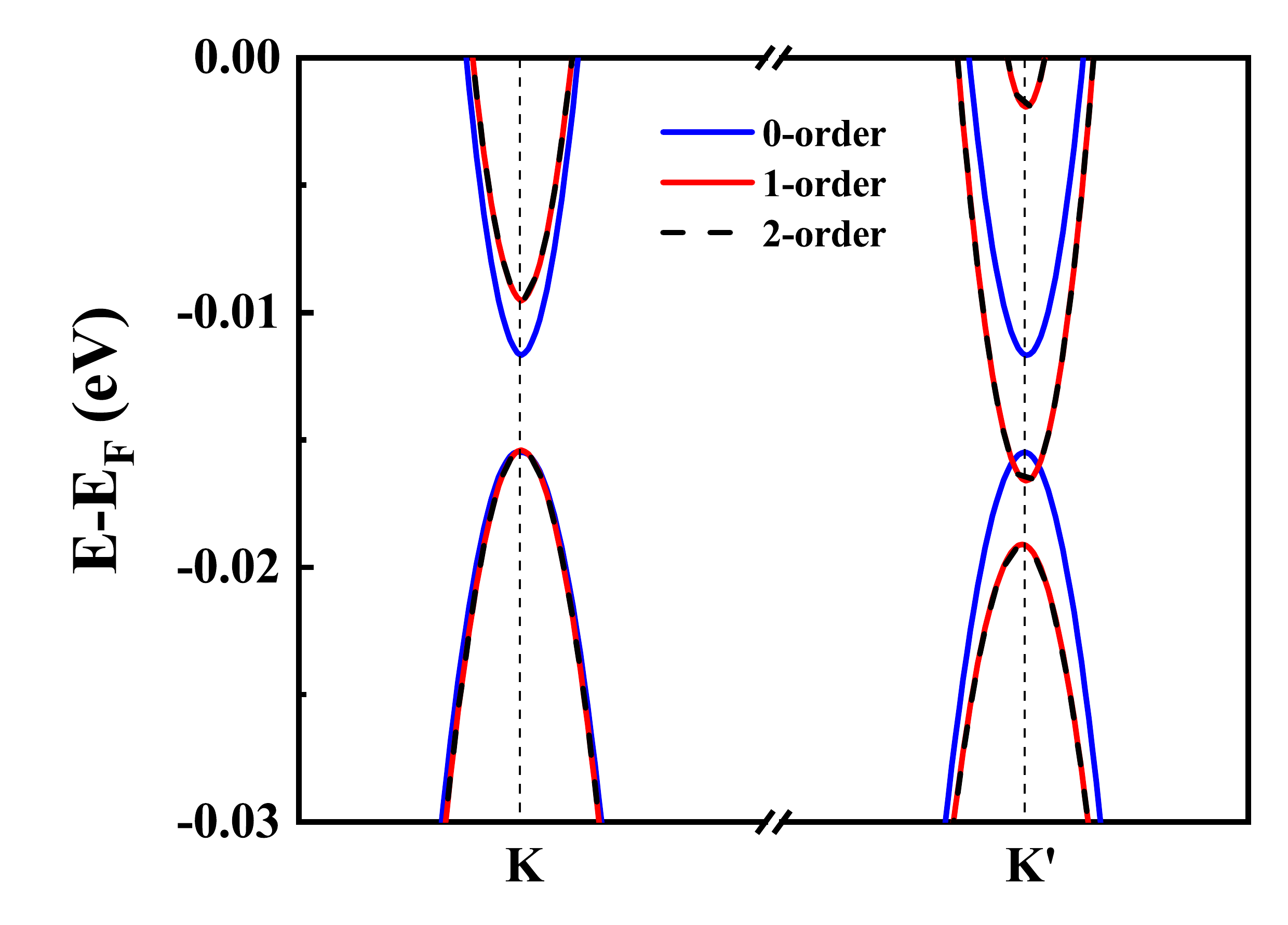}
    \caption{The comparison of Floquet band structures among different truncated orders q=0, 1, 2. We set $eA/\hbar=0.075\rm{\AA^{-1}}$.}
   \label{Fig. S6}
\end{figure}

\section{Optically switchable topological properties dependent on handedness}

Herein, we give the evolution of topological phases subjected to light irradiation of left- and right- handed CPL.
As shown in Figs. \ref {Fig. S7}(a)-(d), with increasing light intensity of left-handed CPL, the band gap of $K$ valley spin-up states first closes and then reopens; that is, only spin-down states of $K'$ valley preserve the inverted band topology, resulting in a VQSH-to-VQAH topological phase transition occurs in $K$ valley. Conversely, as shown in Figs. \ref {Fig. S7}(e)-(f), this topological phase transition of $K'$ valley can be obtained by using the right-handed CPL. To further verify light-driven specific topological phases, as shown in Figs. \ref {Fig. S8} and \ref {Fig. S9}, we calculate evolution of Wannier charge centers (WCCs) and Berry curvature under diffirent helicity of CPL, which indicate helicity of CPL can effectively controll this topological spin-valley filter.

\begin{figure}[H]
    \centering
    \includegraphics[width=17.3cm]{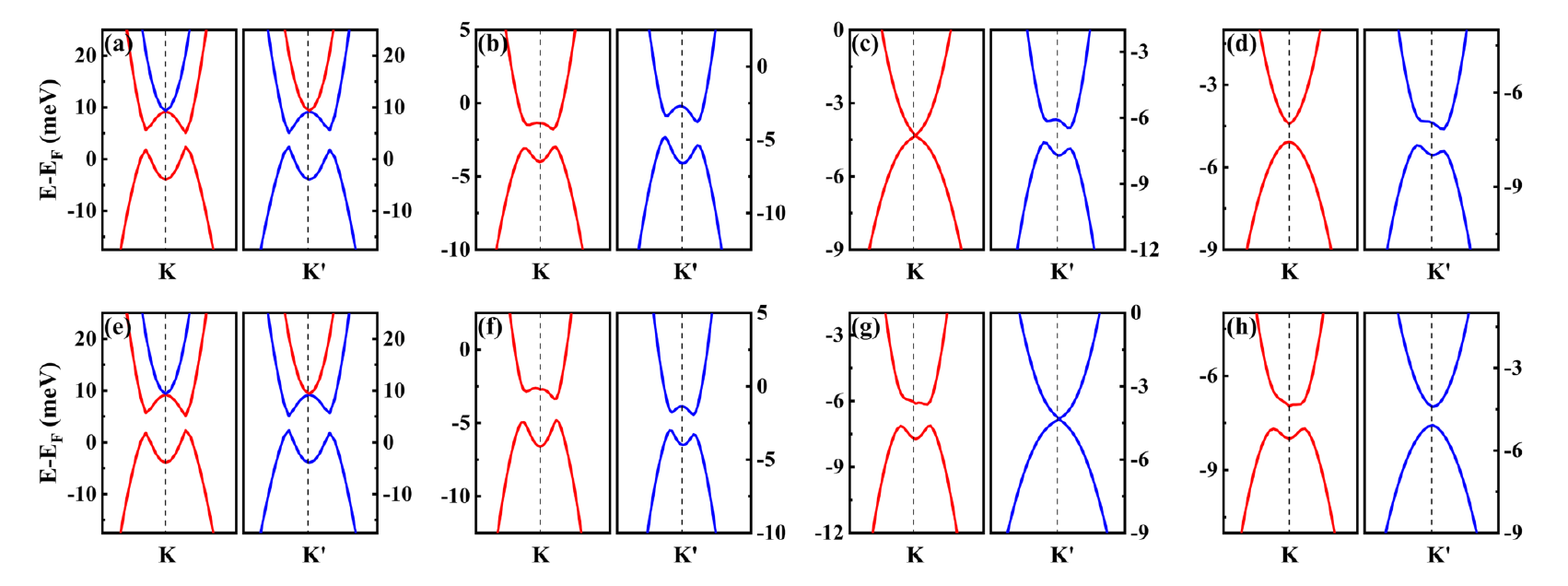}
    \caption{(a)-(d) The evolution of spin-resolved band structures of $H_X^M$ stacking WS$_2$/WTe$_2$ heterobilayer around the $K$ and $K'$ valleys under left-handed CPL with a light intensity $eA/\hbar$ of 0, 0.045, 0.051, and 0.053 {\AA}$^{-1}$. (e)-(h) Likewise under right-handed CPL. The red and blue colors indicate the $z$-component of spin-up and spin-down states, respectively. A perpendicular electric field is fixed at -0.02 V/\AA.}
    \label{Fig. S7}
\end{figure}

\begin{figure}[H]
    \centering
    \includegraphics[width=13cm]{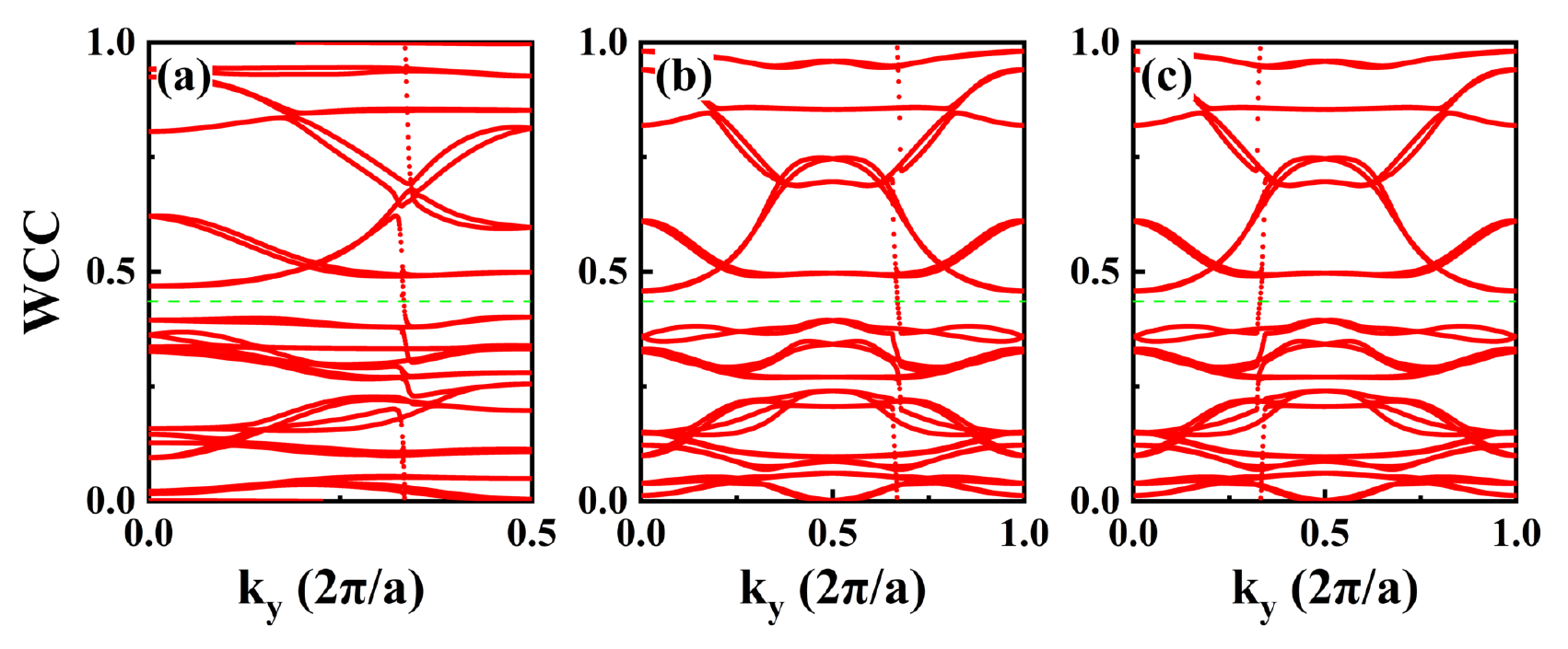}
    \caption{Evolution of Wannier charge centers (WCCs) as a function of $k_y$ (a) without light irradiation and (b) with left-handed or (c) right-handed CPL irradiation ($eA/\hbar=0.053\rm{\AA^{-1}}$).}
    \label{Fig. S8}
\end{figure}

\begin{figure}[H]
    \centering
    \includegraphics[width=17.3cm]{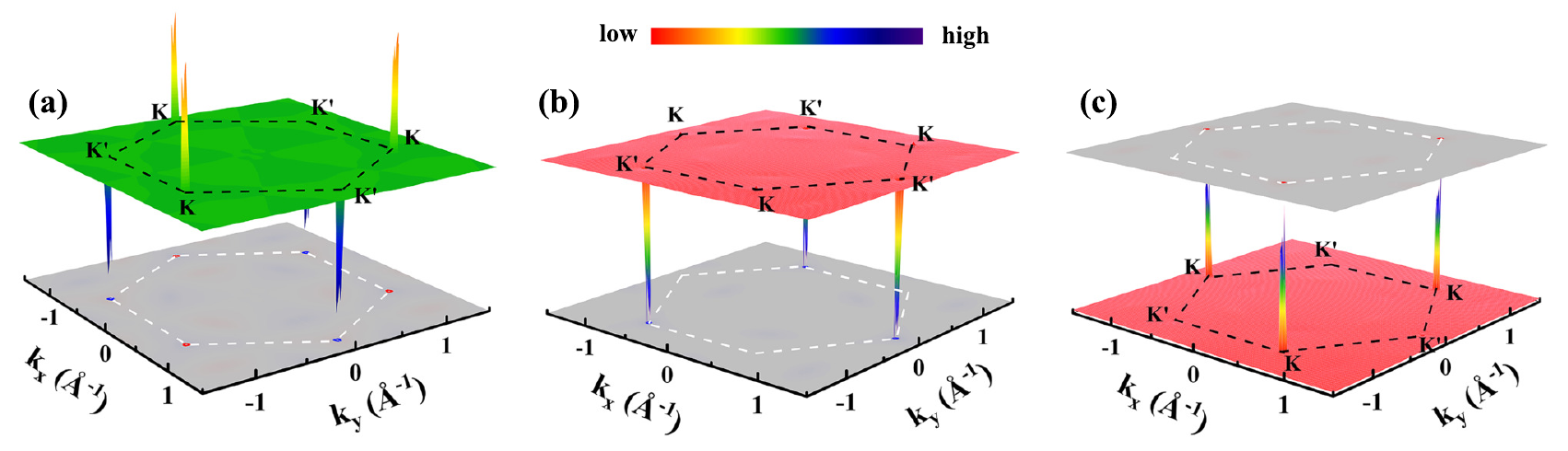}
    \caption{The distribution of the Berry curvature $\Omega_z(\mathbf{k})$ for (a) $\mathcal{T}$-broken (or invariant) VQSH state, (b) VQAH state of $K'$ valley driven by left-handed CPL, and (c) VQAH state of $K$ valley driven by right-handed CPL in the $k_x-k_y$ plane, respectively. The hexagonal Brilllouin zone is marked by dashed lines.}
    \label{Fig. S9}
\end{figure}

%

\vspace{1em}
~~~\\

\end{widetext}

\end{document}